\documentclass[]{aa}
\usepackage{natbib}
\usepackage{graphicx}
\usepackage{color}
\usepackage{txfonts}

\newcommand\bb[1]{\mbox{\boldmath{$#1$}}}
\newcommand\del{\bb{\nabla}}
\newcommand\bcdot{\bb{\cdot}}
\newcommand\btimes{\bb{\times}}
\newcommand\elec{E}

\bibpunct{(}{)}{;}{a}{}{,}

\begin{document}

\title{A High Order Godunov Scheme with Constrained Transport and Adaptive Mesh Refinement for Astrophysical MHD}

\author{ S\'ebastien Fromang \inst{1,2}, Patrick Hennebelle \inst{3}, Romain Teyssier \inst{4}}

\offprints{S.Fromang}

\institute{Department of Applied Mathematics
and Theoretical Physics, University of Cambridge, Centre for
Mathematical Sciences, Wilberforce Road, Cambridge, CB3 0WA, UK \and
Astronomy Unit, Queen Mary, University of London, 
Mile End Road, London E1 4NS, UK \and Laboratoire de radioastronomie
millim\'etrique, UMR 8112 du CNRS,
\'Ecole normale sup\'erieure et Observatoire de Paris, 24 rue Lhomond,
75231 Paris Cedex 05, France \and Service d'Astrophysique,
CEA/DSM/DAPNIA/SAp, Centre d'\'Etudes de Saclay, L'orme des Merisiers,
91191 Gif--sur--Yvette Cedex, France
 \\
\email{S.Fromang@damtp.cam.ac.uk}}
\date{Accepted; Received; in original form;}

\label{firstpage}

\abstract
{}
{ In  this  paper, we  present  a  new  method to  perform  numerical
  simulations  of  astrophysical MHD  flows  using  the Adaptive  Mesh
  Refinement framework and Constrained Transport. }
{ The   algorithm  is  based  on   a  previous  work   in  which  the
  MUSCL--Hancock scheme was used  to evolve the induction equation. In
  this paper, we  detail the extension of this scheme  to the full MHD
  equations and discuss its properties. }
{ Through a series of test problems, we illustrate the performances of
  this  new  code using two different MHD Riemann solvers
  (Lax--Friedrich and Roe) and  the  need  of  the  Adaptive  Mesh Refinement
  capabilities  in some  cases. Finally,  we show  its  versatility by
  applying  it to  two completely  different  astrophysical situations
  well  studied  in the  past years:  the  growth  of the  magnetorotational
  instability in the shearing box and the collapse of magnetized cloud
  cores.}
{ We  have implemented a  new Godunov scheme  to solve the  ideal MHD
  equations in the AMR code RAMSES.  We have shown that it results in
  a  powerful  tool  that  can  be  applied  to  a  great  variety  of
  astrophysical problems, ranging from galaxies formation in the early
  universe to  high resolution studies of molecular  cloud collapse in
  our galaxy.}

\authorrunning{Fromang et al.}
\titlerunning{A High Order Godunov Scheme for Astrophysical MHD}
\maketitle

\section{Introduction}

Developing  efficient  numerical   algorithms  for  the  equations  of
magnetohydrodynamics  (MHD)   is  of  great   astrophysical  interest.
Magnetic  fields are ubiquitous  in a  great variety  of environments.
They are  important components of the  dynamics in such  places as the
early  universe,  the   interstellar  and  intergalactic  medium,  the
environment and interior of stars  and the accretion flow around young
stellar objects.

In the last  few decades, finite differences methods  have been widely
used  in investigations  of a  number of  astrophysical  situations in
which  the  magnetic  field  is  important with  such  codes  as  ZEUS
\citep{stone&norman92a,            stone&norman92b},           NIRVANA
\citep{ziegler&yorke97}        or        the        Pencil        Code
\citep{brandenburg&dobler02} for  example.  Even though,  as expected,
the   numerical    method   breaks   down    in   some   circumstances
\citep{falle02}, a  considerable amount of progress have  been made in
our understanding  of MHD  in astrophysics. A  few attempts  have also
been made to  try to extend the Smoothed  Particle Hydrodynamics (SPH)
method    to   MHD    \citep{phillips&monaghan85,   price&monaghan04a,
price&monaghan04b}.   At the  moment, it  is not  clear,  however, how
efficient the resulting codes will prove to be in the future.

In  the last  few years,  several attempts  have been  made to  try to
extend  the   standard  Godunov  approach   \citep{toro97},  initially
designed to solve the Euler equations, to MHD.  In
addition to the accurate description of new waves that are peculiar to
MHD  (Alfv\'en  waves, the  slow  and fast  modes),  one  of the  most
dramatic challenge in  the development of such schemes  comes from the
solenoidality  constraint, which  states  that the  divergence of  the
magnetic  field has  to  vanish  everywhere at  all  times. The  first
algorithms  that  attempted  to  solve  this  problem  kept  the  cell
centering strategy of the  standard Godunov approach. They used either
a     ``divergence     cleaning''     step    (see     for     example
\citeauthor{Brackbill80}  \citeyear{Brackbill80} or \citeauthor{ryu98}
\citeyear{ryu98}),  or  various reformulations  of  the MHD  equations
including    additional     divergence-waves    \citep{powell99}    or
divergence-damping terms \citep{Dedner02} to enforce the solenoidality
constraint.  A  novel  cell-centered  MHD  scheme  has  been  recently
developed by \cite{Crockett05} that  combines most of these ideas into
one  single  algorithm. Alternative  approach  used the  ``staggered''
discretisation of the grid commonly used in ``ZEUS--like'' codes along
with  the  more   geometrical  Constrained  Transport  (CT)  algorithm
\citep{evans&hawley88}.   This   is    for   example   the   case   of
\citet{balsara&spicer99},              \citet{toth00}              and
\citeauthor{londrillo&delzanna00}     (\citeyear{londrillo&delzanna00},
\citeyear{londrillo&delzanna04}).    \citet{gardiner&stone05a}    also
explored the  possibility of combining  the CT algorithm with  the PPM
scheme in the new code ATHENA.

Recently,  we  proposed   to  extend  the  well--known  MUSCL--Hancock
algorithm originally designed for  the Euler equation to the induction
equation \citep{teyssieretal06}. We showed  that three variants of our
scheme have  good performances.  Two are compatible  with the Adaptive
Mesh    Refinement    (AMR)    algorithm   implemented    in    RAMSES
\citep{teyssier02}.  This first part was limited to the induction
equation, and could  only be applied to situations  where the magnetic
field does  not affect the flow.   This is  enough, however,  to
capture the physics  of fast
dynamos,  especially with  the help  of the  AMR. Here  we  extend our
approach to the full set of MHD equations and implement it in RAMSES.

The plan of  the paper is as follows:  in section~\ref{num method}, we
present  the details  of the  numerical algorithm.  The  discussion is
based on our earlier  work \citep{teyssieretal06}, where the technical
details of the scheme are presented. In section~\ref{test section 1D}
and \ref{test section 2D}, we
illustrate  the properties  of the  code on  standard 1D  and  2D test
problems. In section~\ref{astro  appli}, it is used to  study a few 3D
flows   of    astrophysical   significance:   the    growth   of   the
magnetorotational instability  in accretion disks and  the collapse of
magnetized cloud  cores. Finally, we  summarise the properties  of the
code    and     highlight    future    possible     developments    in
section~\ref{conclusion}.

\section{The numerical method}
\label{num method}

\subsection{Equations and notations}
\label{eq&not}

The  equations we seek  to solve  are the  usual MHD  equations.  When
written in conservative form, they read:
\begin{eqnarray}
\frac{\partial \rho}{\partial t} + \del \bcdot (\rho \bb{v})  =  0 \, , \\
\frac{\partial \rho \bb{v}}{\partial t} + \del \bcdot (\rho\bb{v}\bb{v} - \bb{B}\bb{B}) + \del P_{tot} =0 \, , \\
\frac{\partial E}{\partial t} + \del \bcdot \left[ (E+P_{tot})\bb{v}-\bb{B}(\bb{B} \bcdot \bb{v}) \right] =0 \, , \\
\frac{\partial \bb{B}}{\partial t} + \del \bcdot (\bb{v}\bb{B} - \bb{B}\bb{v})  =  0 \, .
\label{mhd equations}
\end{eqnarray}
Here, $\rho$ is the fluid  density, $\bb{v}$ its velocity and $\bb{B}$
is the  magnetic field. $P_{tot}$  stands for the total  pressure, the
sum of the thermal pressure $P$ and the magnetic pressure:
\begin{equation}
P_{tot}=P+\frac{\bb{B}\bcdot\bb{B}}{2} \, ,
\label{ptot}
\end{equation}
and $E$ is the total energy of the fluid
\begin{equation}
E=\epsilon+\rho\frac{\bb{v}\bcdot\bb{v}}{2}+\frac{\bb{B}\bcdot\bb{B}}{2} \, ,
\label{etot}
\end{equation}
where $\epsilon$ denotes the internal energy. Unless otherwise stated,
we will  assume throughout  this paper that  the equation of  state is
that of a perfect gas, in which case $P=(\gamma - 1) \epsilon$.

As  discussed in the  introduction, this  set of  equations has  to be
completed by the solenoidal constraint, to be satisfied at all times:
\begin{equation}
\del \bcdot \bb{B} = 0
\end{equation}
As in \citet{teyssieretal06}, we will use throughout this paper the CT
scheme  \citep{evans&hawley88}  to   ensure  that  this  condition  is
fulfilled  to  machine--roundoff  precision.  It  simply  consists  in
writing the induction equation in integral form:
\begin{equation}
\frac{\partial \Phi_s}{\partial t}=\frac{\partial}{\partial t} \int\!\!\!\int \bb{B} \bcdot \bb{dS} = \oint \bb{\elec} \bcdot \bb{dl} \, ,
\label{induction}
\end{equation}
where  $\bb{\elec}$ is  the  electric field  defined  by the  relation
$\bb{\elec}=\bb{v}  \btimes   \bb{B}$.  While  all   the  hydrodynamic
variables  (density, velocities,  total  energy) are  located at  cell
centers, this approach requires the magnetic field components to lie on
the cell  faces. The grid structure  that results is  described in the
following section.

\subsection{The staggered mesh}

In  the  followings,  we  describe  our  scheme  using  3  dimensional
coordinates $x$,  $y$ and $z$. The physical  variables are discretized
on a standard  3D Cartesian grid.  The center of  each cell is located
at the position $(x_i,y_j,z_k)$.  In a given cell, faces normal to the
x--direction have  coordinates $x=x_{i \pm  1/2}$ and cover  a surface
element       defined       by       $y_{j-1/2}<y<y_{j+1/2}$       and
$z_{k-1/2}<z<z_{k+1/2}$. The coordinates of the other faces, normal to
the $y$ and $z$ direction, can be similarly defined.

As  for the  Euler equations,  the hydrodynamical  variables (density,
momentum,  energy)   are  volume--averaged  over  a   cell  and  the
discretized values are defined at the cell center. For example:
\begin{equation}
\rho_{i,j,k}=\frac{1}{\Delta x \Delta y \Delta z}
\int_{x_{i-1/2}}^{x_{i+1/2}}
\int_{y_{j-1/2}}^{y_{j+1/2}}
\int_{z_{k-1/2}}^{z_{k+1/2}}
 \rho(x',y',z') dx'dy'dz'
\end{equation}
Because  of   the  staggered  mesh   representation,  magnetic  fields
components are surface--averaged over the cell face to give:
\begin{equation}
B_{x,i-1/2,j,k}=\frac{1}{\Delta y \Delta z}
\int_{y_{j-1/2}}^{y_{j+1/2}} 
\int_{z_{k-1/2}}^{z_{k+1/2}} 
B_x(x_{i-1/2},y',z') dy'dz'
\end{equation}
Here $\Delta x$, $\Delta y$ and $\Delta z$  stand for the Cartesian mesh
size in each direction.

\subsection{The Euler system}
\label{euler section}

As  outlined  in   \citet{londrillo&delzanna00},  the  system  of  MHD
equations  written  in  section~\ref{eq&not}  can  be  broken  in  two
sub-systems.    The  first   involved  the   time  evolution   of  the
cell--centered, volume--averaged  variables and is  reminiscent of the
standard Euler equations, which includes mass, momentum
and  energy  conservation.  This  set  of  equations,  quite
naturally called the  ``Euler system'', can be written  in vectorial
form
\begin{equation}
\frac{\partial \bb{U}}{\partial t}
+\frac{\partial \bb{F}}{\partial x}
+\frac{\partial \bb{G}}{\partial y}
+\frac{\partial \bb{H}}{\partial z}=0 \, ,
\label{compact form}
\end{equation}
where
\begin{equation}
\bb{U}=(\rho,\rho v_x,\rho v_y,\rho v_z, E)^{T}
\end{equation}
and the flux function $\bb{F}$ is given by
\begin{equation}
\bb{F}=\left( \begin{array}{c} \rho v_x \\ 
\rho v_x^2 +P_{tot} - B_x^2 \\ 
\rho v_x v_y - B_x B_y \\ 
\rho v_x v_z - B_x B_z \\ 
(E+P_{tot})v_x - B_x(\bb{B} \bcdot \bb{v}) \end{array} \right) \, .
\end{equation}
The  expression for  $\bb{G}$ for  $\bb{H}$ are  completely symmetric.
Integrating  in  space over  a  cell and  in  time  between $t^n$  and
$t^{n+1}$, equation~(\ref{compact form}) writes:
\begin{eqnarray}
\nonumber
\frac{\bb{U}^{n+1}_{i,j,k}-\bb{U}^n_{i,j,k}}{\Delta t}
&+&\frac{\bb{F}^{n+1/2}_{i+1/2,j,k}-\bb{F}^{n+1/2}_{i-1/2,j,k}}{\Delta x}\\
&+&\frac{\bb{G}^{n+1/2}_{i,j+1/2,k}-\bb{G}^{n+1/2}_{i,j-1/2,k}}{\Delta y}\\
\nonumber
&+&\frac{\bb{H}^{n+1/2}_{i,j,k+1/2}-\bb{H}^{n+1/2}_{i,j,k-1/2}}{\Delta z}=0
\label{discretize hydro}
\end{eqnarray}
where superscripts  n and n+1  refer respectively to  time coordinates
$t^n$  and  $t^{n+1}$.   $\bb{U}^n_i$  and  $\bb{U}^{n+1}_i$  are  the
volume--averaged variables at time $t^n$ and $t^{n+1}$. The time-- and
surface--averaged fluxes are defined by
\begin{eqnarray}
\nonumber
\bb{F}_{i+1/2,j,k}^{n+1/2} & = & \\
 \frac{1}{\Delta t\Delta y \Delta z}
&  \int_{t^n}^{t^{n+1}}  
\int_{y_{j-1/2}}^{y_{j+1/2}}
\int_{z_{k-1/2}}^{z_{k+1/2}} & 
\bb{F}(x_{i+1/2},y',z',t')dy'dz'dt'
\end{eqnarray}
where  $\Delta  t=t^{n+1}-t^n$.   Similar  time  and  surface--average
quantities  are written for  the fluxes  $G^{n+1/2}_{i,j+1/2,k}$ and
$H^{n+1/2}_{i,j,k-1/2}$   appearing   in   equation~(\ref{discretize
hydro}).

In  this paper,  we  intend to  extend  the well-known  MUSCL--Hancock
scheme \citep{vanleer77,toro97}  to the equations of  ideal MHD.  When
applied to the Euler  equations, this method performs the conservative
update  of the  volume--average  variables $\bb{U}$  in  two steps:  a
predictor  step  and a  corrector  step.  In  the former,  the  vector
$\bb{U}$ is computed  at the half time step $t^{n+1/2}  = t^n + \Delta
t/2$ using a Taylor expansion  of the underlying hyperbolic system. It
is also spatially reconstructed from the cell center to the cell faces
using a  piecewise linear reconstruction based on  TVD slope limiters.
In     the    predictor    step,     the    fluxes     appearing    in
equation~(\ref{discretize  hydro})  are  evaluated  by  solving  a  1D
Riemann problem between the  two (left and right) reconstructed states
at each cell interface.

\subsubsection{Cell-centered TVD slopes}

The  first step  in the  MUSCL  approach is  the computation  of
finite--difference  approximation of  the spatial  derivatives  of all
cell--centered  quantities. As  usually  done in  higher order  finite
volume  schemes,  spatial  derivatives  are approximated  using  slope
limiters, in  order to  obtain positivity preserving,  non oscillatory
solutions.  Except  for the final  conservative update, we  always use
the primitive variables  $\bb{W} = (\rho, v_x, v_y,  v_z, P)^T$ in all
intermediate  calculations.   In addition  to  these 5  cell--centered
variables,   we  need  to   define  volume--averaged   magnetic  field
components.    We  use  for   that  purpose   the  average   of  their
corresponding face--centered components
\begin{equation}
\label{vol_avg}
B^n_{x,i,j,k}=\frac{1}{2} \left(
B_{x,i-1/2,j,k}^{n}+B_{x,i+1/2,j,k}^{n} \right) \, ,
\end{equation}
and likewise for $B^n_{y,i,j,k}$ and $B^n_{z,i,j,k}$.
We finally augment the vector  $\bb{W}^n$ with these 3 new components.
We use in our current implementation two standard slope limiters (used
in many fluid  dynamics codes), the MinMod slope and the MonCen
(Monotonized Central) slope \citep{toro97}. The MinMod  limiter is
more diffusive than the MonCen  limiter, so we
use the latter  for most applications. On the
other hand, the MinMod limiter is known to ensure the positivity of
the solution in multiple space dimensions.   In difficult cases,  we
therefore  switch to  the MinMod slope limiter.

\subsubsection{The Euler predictor step}

The standard MUSCL methodology can be applied to the Euler sub--system,
using  the  previously  defined  cell-centered  vector  $\bb{W}$.  The
solution is advanced  in time up to $t^{n+1/2}$ using a Taylor
expansion of the Euler system in {\it non--conservative} form based on
the previously computed TVD slopes
\begin{eqnarray}
\nonumber
\bb{W}^{n+1/2}_{i,j,k} & = & \bb{W}^n_{i,j,k}
- \bb{A_x}^n_{i,j,k}
\left( \frac{\partial \bb{W}}{\partial x}\right)^n_{i,j,k} \Delta t/2 \\
 & -&  \bb{A_y}^n_{i,j,k}
\left( \frac{\partial \bb{W}}{\partial y}\right)^n_{i,j,k} \Delta t/2
- \bb{A_z}^n_{i,j,k}
\left( \frac{\partial \bb{W}}{\partial z}\right)^n_{i,j,k} \Delta t/2
\end{eqnarray}
where the  matrix $\bb{A_x}$  (resp.$\bb{A_y}$ and $\bb{A_z}$)  is the
Jacobian  matrix of  the  flux function  in  the x  (resp.   y and  z)
direction,     evaluated     using     the    cell--averaged     state
$\bb{W}^n_{i,j,k}$.   At this stage,  we have  cell-centered predicted
states   at    time   $t^{n+1/2}$   for   the    5   Euler   variables
$\bb{W}^{n+1/2}_{i,j,k} = (\rho, v_x, v_y, v_z, P)^T$.  Face--centered
predicted values  for the magnetic field components  are also computed
using  the  method  described  in  details  in  section~\ref{induction
section}. We  compute the  predicted cell--centered components  of the
magnetic field using the average of their corresponding face--centered
values.  We  finally augment the  vector $\bb{W}^{n+1/2}_{i,j,k}$ with
these 3 new cell--centered predicted variables.

\subsubsection{The Euler corrector step with 1D Riemann solvers}

Using the  TVD slopes computed at time $t^n$,  we reconstruct
the  primitive  variables  at  each  cell-interface,  except  for  the
longitudinal magnetic  field component, since its  predicted value has
been    already    computed    at    the   correct    location    (see
Sect.~\ref{induction  section}).  For example,  at the  two interfaces
perpendicular   to  the   x--axis,   we  obtain   the  two   following
reconstructed states
\begin{equation}
\bb{W}^{n+1/2,L}_{i+1/2,j,k} = \bb{W}^{n+1/2}_{i,j,k} + \left(  \frac{\partial
\bb{W}}{\partial x}\right)^n_{i,j,k} \Delta x/2
\end{equation}
\begin{equation}
\bb{W}^{n+1/2,R}_{i-1/2,j,k} = \bb{W}^{n+1/2}_{i,j,k}
- \left( \frac{\partial \bb{W}}{\partial x}\right)^n_{i,j,k} \Delta x/2 
\end{equation}
These  states will  be used  as input  states for  1D  Riemann problems
perpendicular to  each interfaces. Note  that for these 1D  MHD Riemann
problems, left  and right states  are defined using only  7 variables,
namely  the 5  Euler variables  and  the 2  magnetic field  transverse
components,  thus  the  name  ``seven waves  Riemann  solvers''. As
far as the Riemann solver is concerned, the longitudinal component of
the magnetic field is assumed  to be  constant in
time and  space, in order  to enforce the solenoidality  constraint in
one  space  dimension. This  constant  value  is  taken equal  to  the
predicted     value    at     time     $t^{n+1/2}$,     namely
$B^{n+1/2}_{x,i+1/2,j,k}$    (resp.    $B^{n+1/2}_{y,i,j+1/2,k}$   and
$B^{n+1/2}_{z,i,j,k+1/2}$)  for the interface  perpendicular to  the x
(resp. y and  z) axis. The output of these 1D  Riemann solvers are the
time   and   surface   averaged   fluxes   at   the   same   interface
$F^{n+1/2}_{i+1/2,j,k}$     (resp.      $G^{n+1/2}_{i,j+1/2,k}$    and
$H^{n+1/2}_{i,j,k+1/2}$).  In  our current implementation,  we use two
different Riemann  solvers, namely a simple, local  Lax Friedrich (LLF)
solver, for which the flux is given by
\begin{equation}
\bb{F}_{LLF}(\bb{W}_L,\bb{W}_R)=\frac{1}{2}\left(\bb{F}_L+\bb{F}_R\right)
-\frac{1}{2} \max_{\alpha=1,7} | \lambda_{\alpha} | \left(\bb{U}_R - \bb{U}_L  \right)
\end{equation}
and  the  MHD  Roe  solver  described  in  \cite{cargo&gallice97}  and
developped by \cite{gardiner&stone05a}, for which the flux can be written as
\begin{eqnarray}
\nonumber
\bb{F}_{Roe}(\bb{W}_L,\bb{W}_R)& = &\frac{1}{2}\left(\bb{F}_L+\bb{F}_R\right)\\
& -& \frac{1}{2} \sum_{\alpha=1,7} \bb{R}_{\alpha} |\lambda_{\alpha}| \bb{L_{\alpha}}
\cdot \left(\bb{U}_R - \bb{U}_L  \right)
\label{roe solver}
\end{eqnarray}
where  $\bb{U}_L$ and
$\bb{U}_R$ are the conservative state on each sides of the interface,
$\bb{F}_L$ and $\bb{F}_R$ the associated fluxes, $\bb{R}_{\alpha}$
and $\bb{L}_{\alpha}$  are respectively  the right and left
eigenmatrices  of the Roe matrix and $\lambda_{\alpha}$ its 
eigenvalues (wave speeds).

\subsection{The induction system}
\label{induction section}

To form  the full set of MHD  equations, equation~(\ref{compact form})
has  to  be completed  by  the  induction  equation, called  here  the
induction sub--system. As for the Euler system, the induction equation
can be written in  conservative  form by  a  straightforward integration  in
space--time
\begin{eqnarray}
\label{discretize mhd}
\frac{B^{n+1}_{x,i-1/2,j,k}-B^n_{x,i-1/2,j,k}}{\Delta t}
& - &\frac{E^{n+1/2}_{z,i-1/2,j+1/2,k}-E^{n+1/2}_{z,i-1/2,j-1/2,k}}{\Delta y}\\
& + &\frac{E^{n+1/2}_{y,i-1/2,j,k+1/2}-E^{n+1/2}_{y,i-1/2,j,k-1/2}}{\Delta z}=0 
  \, ,  \nonumber       
\end{eqnarray}
with  similar expressions for  $B^{n+1}_{y}$ and  $B^{n+1}_{z}$.  Here
conventions  are  similar  to  the  ones  used  in  section~\ref{euler
section}   above,  except   that  one   defines  now   a   time--  and
edge--averaged electromotive force (EMF) as
\begin{eqnarray}
E^{n+1/2}_{z,i-1/2,j-1/2,k} = & \nonumber \\
\frac{1}{\Delta t \Delta z} 
\int_{t_n}^{t_{n+1}} 
\int_{z_{k-1/2}}^{z_{k+1/2}} & 
\elec_z(x_{i-1/2},y_{j-1/2},z',t')dz'dt' \, ,
\end{eqnarray}
and similar   expressions   can  be   derived   for  $E^{n+1/2}_{x}$   and
$E^{n+1/2}_{y}$.  As for the Euler system, the numerical evaluation or
the  EMF  proceeds  in two  steps:  a  predictor  step followed  by  a
corrector step. The MUSCL methodology can be extended to the induction
system   and    this   extension   was    extensively   discussed   in
\citet{teyssieretal06}.  We recall here only the basic ingredients.

\subsubsection{Face-centered TVD slopes}

In  order  to obtain  a  second--order  accurate and  non--oscillatory
solution, we need to use  spatial reconstruction of the magnetic field
components based  on TVD slope  limiters.  The main  difference arises
because of  the finite--surface representation of  the magnetic field.
Indeed, we need a piecewise  linear representation of $B_x$ within the
y--z plane. For  that purpose, we use the same  TVD slopes (MinMod and
MonCen) as  above, using  the face--averaged value  of the  3 magnetic
field components at time $t^n$. For $B_x$, we need to compute only the
2  transverse slopes  $\partial B_x  / \partial  y$ and  $\partial B_x
/\partial z$. A similar property holds for $B_y$ and $B_z$.

\subsubsection{The induction predictor step}
\label{pred step}

Various methods to perform the predictor step for the induction system
were  recently explored  by \citet{teyssieretal06}  for  the kinematic
case.  These  methods were referred  to as Runge--Kutta,  U--MUSCL and
C--MUSCL.   The extension  of the  first two  to the  full set  of MHD
equations, while  possible, is computationally  expensive because they
require to solve one (U--MUSCL) or two (Runge--Kutta) Riemann problems
in  the   predictive  step.   Moreover,  the  large   stencil  of  the
Runge--Kutta  scheme  is  not  compatible  with  the  compact  stencil
required  by our  tree-based  AMR implementation.   For these  various
reasons, we decide to use only  the C--MUSCL scheme in our current MHD
application.    It    combines   the   nice    properties   of   being
computationnally efficient  and compatible with  the AMR requirements.
The price to pay is a reduced stability range for the time step, since
the Courant  factor has to  be less than $2/(\sqrt{2}+1)$,  instead of
$1$ for the other schemes (see \citeauthor{teyssieretal06}
\citeyear{teyssieretal06} for details) .

The purpose of the predictive  step is to advance the solution between
$t^n$ and $t^{n+1/2}$ using the CT algorithm. For that, EMF need to be
spatially  interpolated on  cell  edges  at time  $t^n$.  The idea  of
C--MUSCL is  to do  it by simple  arithmetic averages of  the magnetic
field    and    velocity   components.    For    example,   the    EMF
$E_{z,i-1/2,j-1/2,k}^{n}$ is calculated by:
\begin{equation}
E_{z,i-1/2,j-1/2,k}^{n}=\bar{v_x}\bar{B_y}-\bar{v_y}\bar{B_x} \, ,
\end{equation}
with
\begin{eqnarray}
\bar{v_x} &=& \frac{1}{4} (v^n_{x,i,j,k}+v^n_{x,i-1,j,k}+v^n_{x,i,j-1,k}+v^n_{x,i-1,j-1,k}) \, , \\
\bar{v_y} &=& \frac{1}{4} (v^n_{y,i,j,k}+v^n_{y,i-1,j,k}+v^n_{y,i,j-1,k}+v^n_{y,i-1,j-1,k}) \, , \\
\bar{B_x} &=& \frac{1}{2} (B^n_{x,i-1/2,j,k}+B^n_{x,i-1/2,j-1,k}) \, , \\
\bar{B_y} &=& \frac{1}{2} (B^n_{y,i,j-1/2,k}+B^n_{y,i-1,j-1/2,k})\, .
\end{eqnarray}
This  spatial reconstruction  is second  order in  space,  altough TVD
slopes    have   not   been    used   at    that   time.     The   EMF
$E_{x,i,j-1/2,k-1/2}^n$,          $E_{y,i-1/2,j,k-1/2}^n$          and
$E_{z,i-1/2,j-1/2,k}^n$ are  then used to update  the solution between
$t_n$ and $t_{n+1/2}$ using  the CT algorithm (see Eq.~\ref{discretize
mhd}).   Because  only  one  EMF  is calculated  per  cell  edge,  the
predicted face--centered  magnetic field ($B_x^{n+1/2}$, $B_y^{n+1/2}$
and  $B_z^{n+1/2}$) satisfies  the  solenoidality constraint  exactly.
The properties of C--MUSCL and  their comparison with the U--MUSCL and
Runge--Kutta schemes  are described in  \citet{teyssieretal06}. It was
found that  C--MUSCL behaves essentially similarly to  these two other
schemes,  with a  lower  computational cost  and  a slightly  stronger
Courant condition.

\subsubsection{The induction corrector step with 2D Riemann solvers}
\label{2D Riemann solver}

As described above,  after the predictor step, we  have obtained the 5
cell--centered Euler  variables $\bb{W}^{n+1/2}_{i,j,k} =  (\rho, v_x,
v_y,  v_z, P)^T$ and  the 3  face--centered magnetic  field components
$B_x^{n+1/2}$, $B_y^{n+1/2}$ and $B_z^{n+1/2}$.  Using
equation~(\ref{vol_avg}) at time $t^{n+1/2}$, we have also obtained
the 3 predicted cell--centered  components of the magnetic field. We
finally augment   the  vector  $\bb{W}^{n+1/2}_{i,j,k}$   with  these
3  new cell--centered predicted variables.
 
Following  again the  MUSCL methodology,  we now  need  to reconstruct
complete MHD  states at each cell--edge,  in order to  compute the EMF
for the final conservative update of the magnetic field
components. This reconstruction will produce 4 different states in the
4 cells adjacent to the edge. For obvious reasons, these 4
states, separated by 4 boundaries  (labelled N for North, S for South,
W for West and E for  East) will be labelled in clockwise order by  NE, SE, SW
and NW. 

We first reconstruct the cell--centered state to the cell edges using
\begin{equation}
\bb{W}^{n+1/2,NE}_{i-1/2,j-1/2,k} = \bb{W}^{n+1/2}_{i,j,k} - \left(  \frac{\partial
\bb{W}}{\partial x}\right)^n_{i,j,k} \frac{\Delta x}{2}
- \left(  \frac{\partial
\bb{W}}{\partial y}\right)^n_{i,j,k} \frac{\Delta y}{2}
\end{equation}
and similar relations that defines
$\bb{W}^{n+1/2,SE}_{i-1/2,j-1/2,k}$,
$\bb{W}^{n+1/2,SW}_{i-1/2,j-1/2,k}$ and $\bb{W}^{n+1/2,NW}_{i-1/2,j-1/2,k}$.
Since the 2 longitudinal  magnetic field components ($B_x^{n+1/2}$ and
$B_y^{n+1/2}$) are  already defined at the 4  adjacent interfaces, we
only need the cell--centered transverse component $B_z^{n+1/2}$ in the
above reconstruction. $B_x^{n+1/2}$ and $B_y^{n+1/2}$ are reconstructed
using face--centered TVD slopes as
\begin{eqnarray}
B^{n+1/2,S}_{x,i-1/2,j-1/2,k} = B^{n+1/2}_{x,i-1/2,j-1,k} + \left(  \frac{\partial
B_x}{\partial y}\right)^n_{i-1/2,j-1,k} \frac{\Delta y}{2} \\
B^{n+1/2,W}_{y,i-1/2,j-1/2,k} = B^{n+1/2}_{y,i-1,j-1/2,k} + \left(  \frac{\partial
B_y}{\partial x}\right)^n_{i-1,j-1/2,k} \frac{\Delta x}{2}
\end{eqnarray}
and similar relations defining $B^{n+1/2,N}_{x,i-1/2,j-1/2,k}$ and
$B^{n+1/2,E}_{y,i-1/2,j-1/2,k}$. The four corner states, with  6
variables each, and the 4 longitudinal
magnetic field  components entirely define  a 2D MHD  Riemann problem,
which  satisfies   the  solenoidality   constraint  in  a   2D  sense.
\citet{londrillo&delzanna00} have  shown that the EMF  entering in the
final Contrained  Transport update should be obtained  as the solution
of  this 2D Riemann  problem, in  order to  obtain a  stable numerical
solution,  with a proper  upwinding of  all MHD  waves.  While  a very
simple    exact    solution     exists    in    the    kinetic    case
\citep{teyssieretal06}, designing 2D Riemann  solvers for the full set
of MHD equations is an ambitious task that is beyond the scope of this
paper.  An approximate solution, proposed  by
\citet{balsara&spicer99} and \citet{ziegler04}, is based on averaging the
flux given by the four adjacent 1D Riemann problems. The solution
$E_z^{2D}$ of the 2D Riemann problem writes in that case:
\begin{eqnarray}
\nonumber
E^{2D}_z(\bb{W}^{NE},\bb{W}^{SE},\bb{W}^{SW},\bb{W}^{NW}) &=&  \\
\nonumber
\frac{1}{4} ( E^{1D}_z(\bb{W}^{NW},\bb{W}^{NE}) &+& E^{1D}_z(\bb{W}^{SW},\bb{W}^{SE}) \\
+E^{1D}_z(\bb{W}^{SW},\bb{W}^{NW}) & + &
E^{1D}_z(\bb{W}^{SE},\bb{W}^{NE}) ) \, ,
\end{eqnarray}
where the quantities $E^{1D}_z$ stands for the solution of the 1D
Riemann problems defined by the four states. This solution relies on 4
Riemann solvers per cell edges and turns out
to     be    quite    expensive.      Following    the     ideas    of
\citeauthor{londrillo&delzanna00}     (\citeyear{londrillo&delzanna00},
\citeyear{londrillo&delzanna04}), we exploit the  fact that in our
current  implementation,  we use  only  {\it  linear} Riemann  solvers
(namely Lax--Friedrich and Roe). In this case, the flux can be written
as in equation~\ref{roe solver}. If we now use 2 Roe matrices, one for
each direction, instead of 4, the EMF function can be written as
\begin{eqnarray}
E^{2D}_z(\bb{W}^{NE},\bb{W}^{SE},\bb{W}^{SW},\bb{W}^{NW}) &=& \nonumber  \\
\nonumber
\frac{1}{4} ( E_z(\bb{W}^{NE}) + E_z(\bb{W}^{SE}) &+& E_z(\bb{W}^{SW})+ E_z(\bb{W}^{NW})
) \nonumber  \\
 - \frac{1}{2} \sum_{\alpha=1,7} \bb{R}^x_{\alpha} |\lambda^x_{\alpha}| \bb{L^x_{\alpha}}
& \cdot & \left(\bb{U}_E - \bb{U}_W  \right) \nonumber \\
 + \frac{1}{2} \sum_{\alpha=1,7} \bb{R}^y_{\alpha} |\lambda^y_{\alpha}| \bb{L^y_{\alpha}}
& \cdot & \left(\bb{U}_N - \bb{U}_S  \right)
\end{eqnarray}
where  $\bb{U}_E$, $\bb{U}_W$, $\bb{U}_N$  and $\bb{U}_S$  are averaged
conservative  variables  defined at  the  interfaces  of  the four  1D
Riemann problems. One intersesting property in the above expression is
the explicit  contribution of the 2  diffusive terms coming  from the 2
Roe matrices.   For non--linear  Riemann solvers, such  as HLL,  it is
preferable    to   use   the    4   Riemann    solvers   approximation
\citep{londrillo&delzanna04,ziegler04}.

\subsection{The AMR scheme}

The  AMR  algorithm  used  in   RAMSES  is  described  in  details  in
\citet{teyssier02} and its extension to MHD in \citet{teyssieretal06}.
We briefly recall the main features here.  It is a tree-based AMR code
originally designed for cosmological applications.  The data structure
is a  ``Fully Threaded Tree'' \citep{Khokhlov98}. The  grid is divided
into groups  of 8 cells, called  ``octs'', that share  the same parent
cell.  Each oct  has access to its parent cell  address in memory, but
also  to neighboring  parent cells.   When a  cell is  refined,  it is
called a  ``split'' cell, while in  the opposite case, it  is called a
``leaf'' cell.  The computational domain is always defined as the unit
cube,  which corresponds  in our  terminology  to the  first level  of
refinement in the  hierarchy $\ell = 1$. The  grid is then recursively
refined up to  the minimum level of refinement  $\ell_{min}$, in order
to  build the coarse  grid.  This  coarse grid  is the  base Cartesian
grid,  covering the  whole computational  domain, from  which adaptive
refinement can proceed.  This  base grid is eventually refined further
up to some maximum level of refinement $\ell_{max}$, according to some
user defined refinement  criterion.  When $\ell_{max}=\ell_{min}$, the
computational  grid is  a traditional  Cartesian grid,  for  which the
previous scheme  apply without  any modification.  When  refined cells
are created, however, some issues specific to AMR must be addressed.

\subsubsection{Divergence-free Prolongation Operator}

When a  cell is  refined, eight  new cells (i.e.   a new  ``oct'') are
created for which new  cell--centered variables and new magnetic field
components are needed.   This operation is usually referred  to as the
``prolongation  operator''.   The  traditional  approach relies  on  a
conservative  interpolation  of   the  5  cell--centered  conservative
variables $\bb{U}=(\rho,\rho  v_x,\rho v_y,\rho v_z,  E)^{T}$. For the
face--centered variables, each of the six faces of the parent cell are
split into  4 new fine faces.  Three  new faces, at the  center of the
parent  cell,  are also  split  into  four  new children  faces.   The
resulting magnetic  field components, fine or coarse,  need to satisfy
the divergence-free constraint in integral form.

This critical  step  has   been  solved  by  \cite{balsara01}  and
\cite{toth02}  in  the  CT  framework.   We recommend  both  of  these
articles for  a detailed  description of the  method.  The idea  is to
used slope limiters to interpolate the magnetic field component inside
each  parent face,  in a  flux-conserving way,  and then  to use  a 3D
reconstruction, which  is divergence-free in a local  sense inside the
whole  cell  volume,  in  order  to compute  the  new  magnetic  field
components  for each  central children  faces. In  our case,  the same
slope limiters as in the Godunov scheme (MinMod or MonCen) can be used.

This prolongation operator  is used to estimate the  magnetic field in
newly refined cells,  but also to define a  temporary ``buffer zone'',
two ``ghost cells'' wide, that  set the proper boundary for fine cells
at a coarse-fine level boundary. This is the main reason why compact
stencils are needed for the underlying Godunov scheme.

\subsubsection{Magnetic Flux Corrections}

The other  important step is to  define the reverse  operation, when a
split  cell  is de-refined,  and  becomes  a  leaf cell  again.   This
operation is usually called  the Restriction Operator in the multigrid
terminology.    The  solenoidality  constraint   needs  again   to  be
satisfied,  which translates  into conserving  the magnetic  flux. The
magnetic field  component in  the coarse face  is just  the arithmetic
average of the  4 fine face values. This is  reminiscent of the ``flux
correction        step''       for       the        Euler       system
\citep{berger84,berger89,teyssier02}.

\subsubsection{EMF Corrections}

The  ``EMF  correction  step''  is  more  specific  to  the  induction
equation. For a coarse face which  is adjacent, in any direction, to a
refined  face,  the coarse  EMF  in  the  conservative update  of  the
solution needs  to be  replaced by the  arithmetic average of  the two
fine  EMF vectors.  This  guarantees that  the magnetic  field remains
divergence--free, even at coarse--fine boundaries.

\section{Numerical tests in 1D}
\label{test section 1D}

\subsection{Non--linear Alfv\'en wave test}

\begin{figure}
\begin{center}
\scalebox{0.5}{
\includegraphics[60,50][530,550]{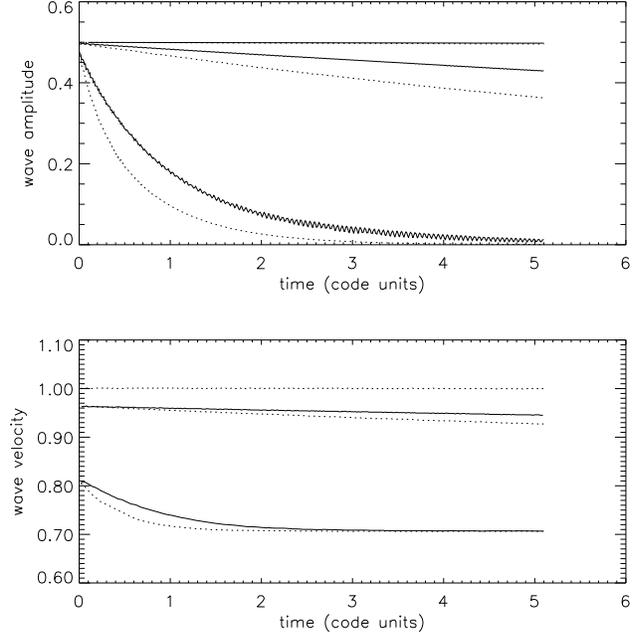}}
\caption{Amplitude ({\it upper panel}) and phase velocity ({\it lower
    panel}) of the  circularly  polarized Alfv\'en  wave as  a
function of  time. The  full lines display  result using a  Roe solver
whereas the  dotted lines show  results obtained with  a Lax-Friedrich
solver. The resolutions are from top to bottom, 100,  30 and 10 grid
points per wavelength}
\label{wave_dissip}
\end{center}
\end{figure}

The first test we present  is the propagation of non-linear circularly
polarised Alfv\'en waves. Such waves,  which are exact solution of the
MHD equations,  propagate in  a gas of  uniform density,  $\rho_0$ and
along a  uniform magnetic field, $B_{0z}$.  They are given  by: $B_x =
B_\perp cos(\omega  t - k  z)$, $B_y =  B_\perp sin(\omega t -  k z)$,
$V_x = V_\perp cos(\omega  t - k z)$, $V_y = V_\perp  sin(\omega t - k
z)$ where $\omega  / k = B_{0z} / \sqrt{4 \pi  \rho_0}$ and $B_\perp /
V_\perp = \sqrt{4 \pi \rho_0}$.

We have  simulated the propagation of  these waves on  a uniform grid,
for  $B_{0z}=\sqrt{4 \pi}$,  $B_\perp=\sqrt{\pi}$,  and $k=0.1$.  This
leads  to  a wave  period  equal to  0.1.  The  agreement between  the
analytical   and  numerical   solutions  depends   on   the  numerical
resolution.  Fig.~\ref{wave_dissip} displays the  wave amplitude ({\it
  upper panel}) and phase velocity ({\it lower panel}) as a function
of time for the Roe and Lax-Friedrich  solvers and different
resolutions,  namely 10,  30 and  100 points  per wavelength.  With 10
grid points  per  wavelength,  the  amplitude  quickly  decays
because of
numerical dissipation  and in about  5 wave periods, the  amplitude of
the waves is only 40$\%$ of  its initial value. With 30 and 100 grid
points 
per wavelength  the agreement is much  better and almost  no decay has
occurred even  after 50 wave periods  in the latter  case. In the
lower panel, the wave velocity is also seen to agree better and better
with its theoretical value, $v_w=1$, as the resolution is increased (note
that for 100 grid points, the wave velocity obtained with the Roe solver is
not represented as it is indistinguishable from the Lax-Friedrich
results). As expected the Roe solver leads to slightly better results
in both cases than the Lax-Friedrich solver.

\subsection{MHD shock tube test}

\begin{figure}
\begin{center}
\includegraphics[scale=0.44]{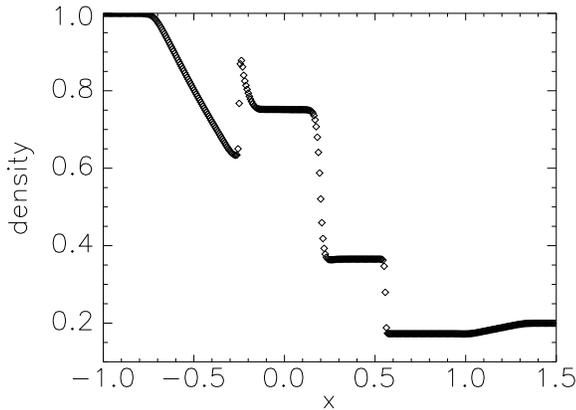}
\caption{Solution  to the  MHD shock  tube  showing the  density as  a
function of x,  obtained with RAMSES when $\alpha=\pi$.  The code uses
$400$ grid points in that case and the Roe Riemann solver.}
\label{tubachoc mhd pi}
\end{center}
\end{figure}
An  interesting application  of the  AMR scheme  is the  study  of the
development of compound waves in  shock tube calculations. It has been
analysed  with finite volume  schemes by  \citet{torrilhon04}, through
the analysis of the MHD shock tube whose initial state is:
\begin{eqnarray}
\bb{W^L}&=&(1,0,0,0,1,1,1,0)^T \\
\bb{W^R}&=&(0.2,0,0,0,0.2,1,\cos \alpha,\sin \alpha)^T
\end{eqnarray}  
where  $\bb{W}=(\rho,v_x,v_y,v_z,P,B_x,B_y,B_z)^T$. When $\alpha=\pi$,
there are two solutions to  the Riemann problem: the first is regular,
which   means    that   it   contains   only    shocks   and   contact
discontinuities. The  second, however, features a  compound wave which
is a composition  of an Alfv\'en and a  slow wave. \citet{torrilhon04}
showed  that finite volume  schemes converge  toward the  second. When
$\alpha$ is different  from $\pi$, the solution is  regular and should
only   contain   shocks    and   contact   discontinuities.   However,
\citet{torrilhon04}  found  that finite  volume  codes  still tend  to
exhibit  the  compound wave  for  low  and  moderate resolutions.  The
solution converges  toward the regular  solution only when  very large
resolutions are used.

Here, we use RAMSES to illustrate how the AMR scheme can help to solve
this  problem.  Figure~\ref{tubachoc mhd  pi}  shows  the density  vs.
position when  $\alpha=\pi$ at time  $t=0.4$. The grid is  composed of
$400$  cells evenly  distributed between  $-1$ and  $1.5$. The  AMR is
switched off in  this first run. The compound  wave is clearly visible
at  $x\simeq-0.25$. The  whole  solution looks  identical to  previous
results   published    in   the   literature    with   similar   codes
\citep{ryu&jones95,cargo&gallice97,londrillo&delzanna00}.

\begin{figure}
\begin{center}
\includegraphics[scale=0.44]{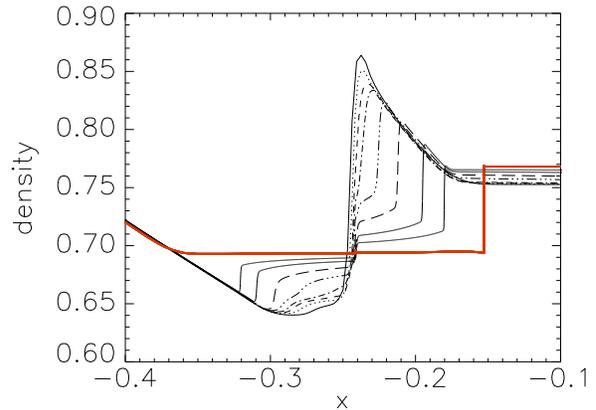}
\caption{Zoom on the  region in which the compound  wave develops when
$\alpha=3$. Black curves  are obtained on a uniform  grid. From top to
bottom (at $x \sim -0.24$),  they correspond to $800$, $1200$, $1600$,
$2000$,   $3000$,   $5000$,    $10000$   and   $20000$   grid   points
respectively. The red curve was  calculated after switching on the AMR
scheme and  using a similar  CPU time as  for the $20000$  grid points
curve.   Its maximum resolution  is equivalent  to using  about $10^6$
cells on  a uniform  grid and show  a dramatic improvement  toward the
regular solution.}
\label{tubachoc mhd zoom}
\end{center}
\end{figure}

\begin{figure}
\begin{center}
\includegraphics[scale=0.44]{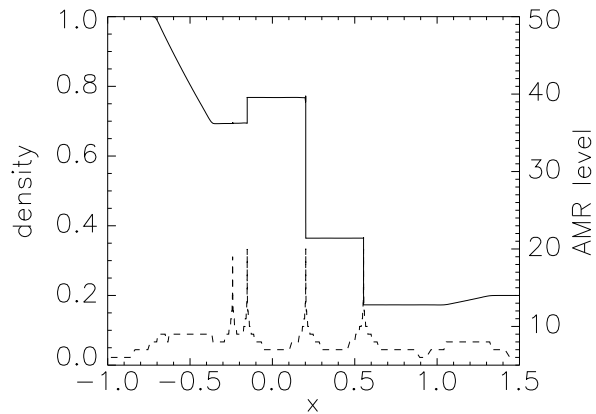}
\caption{Complete solution of the MHD shock tube when $\alpha=3$, with
the AMR scheme turned on. The solid line is a plot of the density as a
function of  position while the dashed  line (whose scale  is given on
the right axis) illustrates the  level of refinement the code uses for
each cell.}
\label{tubachoc mhd amr}
\end{center}
\end{figure}

Taking now $\alpha=3$ and computing the solution of the MHD shock tube
on a  uniform grid, we also  found that the compound  wave remains for
low   resolution  as  described   by  \citet{torrilhon04}.    This  is
illustrated in figure~\ref{tubachoc mhd zoom},  which is a zoom on the
structure  of  the  solution  in  the neighbourhood  of  the  compound
wave. The black lines are computed on a uniform grid and correspond to
increasing resolution. Namely, from top to bottom (at $x \sim -0.24$),
the number of cells are $800$, $1200$, $1600$, $2000$, $3000$, $5000$,
$10000$ and $20000$.  The red line  shows the result of the same model
computed using the AMR scheme  with a refinement strategy based on the
magnitude  of  the gradient  of  all  7  flow variables.   The  finest
resolution  in this  run is  equivalent to  having $10^6$  cells  on a
uniform  grid.   We  found the  result  of  this  model to  be  almost
indistinguishable from the regular solution (remember that this is the
ONLY physical  solution to this problem).  Interestingly,  this is not
the case for  the uniform runs. Even though the  compound wave is seen
to  gradually  disappear  as  the resolution  is  increased,  features
departing  from the  correct  solution are  still  observed even  when
$20000$  grid zones are  used.  This  illustrates the  extremely large
resolution needed to accurately calculate the solution of this problem
and shows  the interest of using  AMR.  Indeed, the AMR  run used only
$10000$ cells for the same equivalent resolution as $10^6$ grid cells,
which corresponds to a gain of about $2500$ in CPU time.

The   complete    solution   of   the   AMR   model    is   shown   on
figure~\ref{tubachoc mhd amr}. As is figure~\ref{tubachoc mhd pi}, the
solid  line represents  the density  as a  function of  position.  The
dashed  line shows the  corresponding refinement  level.  It  scale is
indicated on the  right axis. As expected, the  grid is highly refined
at  the location  of the  shocks and  contact discontinuities.   It is
important to understand that the AMR is not just a fancy tool for this
test, but is actually essential to solve it properly. One might indeed
think that increasing the order  of the numerical scheme would help to
converge    to   the   regular    solution   at    lower   resolution.
\citet{torrilhon&balsara04} actually showed  that the improvement when
using  third or  fourth  order  WENO schemes  is  small. This  is
because the  accuracy of any such  schemes breaks down  to first order
close to  discontinuities, which is precisely where  the compound wave
lies.

\section{Numerical tests in 2D}
\label{test section 2D}
\subsection{Advection of a magnetic loop}

\begin{figure}
\begin{center}
\includegraphics[scale=0.45]{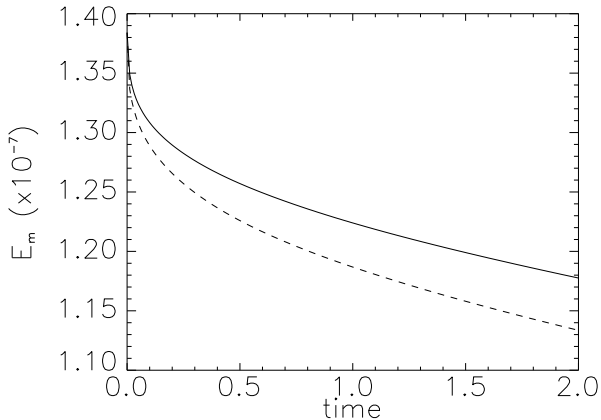}
\caption{Time  history of  the magnetic  energy in  the  magnetic loop
advection  test.  The different  curves  are  obtained  using the  Roe
Riemann solver ({\it solid  line}) and the Lax--Friedrich solver ({\it
dashed line}).}
\label{norm_loop}
\end{center}
\end{figure}
As a first 2D test, we now consider the simple advection of a magnetic
loop that has recently  been proposed by \citet{gardiner&stone05a}. It
simply  consists in  the  evolution of  a  weak magnetic  field in  an
initially uniform  velocity field. Since the thermal  pressure is much
larger  than  the  magnetic   pressure,  the  magnetic  field  can  be
considered  as  a  passive  tracer  advected  in  a  time  independent
flow.   The    initial   setup   is    exactly   the   same    as   in
\citet{gardiner&stone05a}  and \citet{teyssieretal06}.  The velocities
are set up to $v_x=2$, $v_y=1$ and $v_z=0$. The initial magnetic field
is such that $B_z=0$, while the components $B_x$ and $B_y$ are defined
using the z-component of  the potential vector $\bb{A}$ ($\bb{B}= \del
\btimes \bb{A}$):
\begin{equation}
A_z = \left\{
  \begin{array}{cc}
    A_0(R-r) & \rm{for~} r < R \, ,\\
    0 & \rm{otherwise} \, ,
  \end{array}
\right. 
\end{equation}
with $A_0=10^{-3}$, $R=0.3$  and $r=\sqrt{x^2+y^2}$. The computational
domain is  defined as $-1<x<1$ and  $-0.5<y<0.5$. There  are $128$
cells in the  first direction and $64$ in the  second. The solution is
evolved between $t=0$  and $t=2$ and we analyzed  the results obtained
by our scheme using the MonCen slope limiter, comparing explicitly the
Roe solver to the local Lax--Friedrich solver.

A simple  way to evaluate the  efficiency of the scheme  is to compare
the  time history  of  the magnetic  energy  $E_m$.  This  is done  in
figure~\ref{norm_loop}, where $E_m$ is  represented as a function of
time for the  Roe solver ({\it solid line})  and for the Lax-Friedrich
solver ({\it  dashed line}). We first  note that the  results are very
similar  to   those  published  by   \citet{gardiner&stone05a}. This
demonstrates that, using only TVD linear reconstruction, our scheme
provides comparable accuracy to the piecewise parabolic scheme of
\citet{gardiner&stone05a}.  As expected,
the  Lax--Friedrich  solver is  more  dissipative,  while the  results
obtained using  the Roe solver  exactly reproduce the results
obtained in  the pure advection case  \citep{teyssieretal06}, to which
we refer the reader for more details, especially regarding the AMR scheme.

\subsection{The Orszag--Tang vortex}

\begin{figure}
\begin{center}
\includegraphics[scale=0.4]{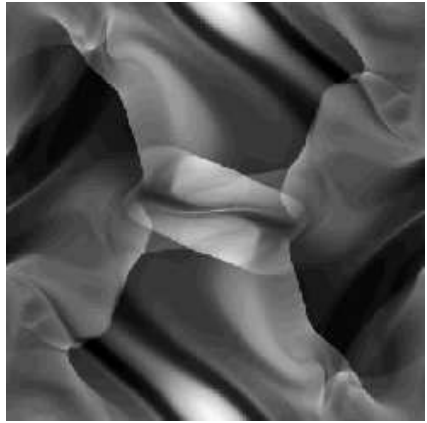}
\caption{Snapshot of  the density at  time $t=0.5$ resulting  from the
Orszag-Tang  test.  The  grid  is  uniform  and  composed  of  $512^2$
cells.  The result is  computed using  the Roe  Riemann solver  and the
C--MUSCL predictive step.}
\label{orszag tang}
\end{center}
\end{figure}

One of the most well--known 2D MHD test is the Orszag--Tang test.  The
initial condition  of the  flow create a  vortex that is  unstable and
quickly breaks down into  turbulence.  Although no analytical solution
is  known for  this test,  it  has been  so widely  documented in  the
literature that it is now very  useful as a first 2D benchmark for 
a code.

The initial state is defined as
\begin{eqnarray}
\rho &=& \gamma P_0 \, , \\
\bb{v} &=& (-\sin 2\pi y,\sin 2\pi x) \, , \\
\bb{B} &=& (- B_0 \sin 2\pi x,B_0 \sin 4\pi y) \, . 
\end{eqnarray}
The different parameters are defined by 
\begin{displaymath}
\gamma=\frac{5}{3} \, , P_0=\frac{5}{12\pi} \textrm{ and } B_0=\frac{1}{\sqrt{4\pi}} \, .
\end{displaymath}
The grid extends from $0$ to $1$ in both directions and we use $512^2$
cells  uniformly  distributed  over  the  computational  domain.   The
solution is evolved between $t=0$ and $t=0.5$ using the Roe solver and
the  MonCen slope  limiter. The  density distribution  in  the $(x,y)$
plane  at the  end of  the simulation  is shown  on figure~\ref{orszag
tang}.  The  agreement between our result and  previous published work
is excellent  \citep{ryu98,londrillo&delzanna00}.  The complex pattern
of interacting waves is perfectly recovered.

\subsection{The current sheet}

\begin{figure*}
\begin{center}
\includegraphics[scale=0.5]{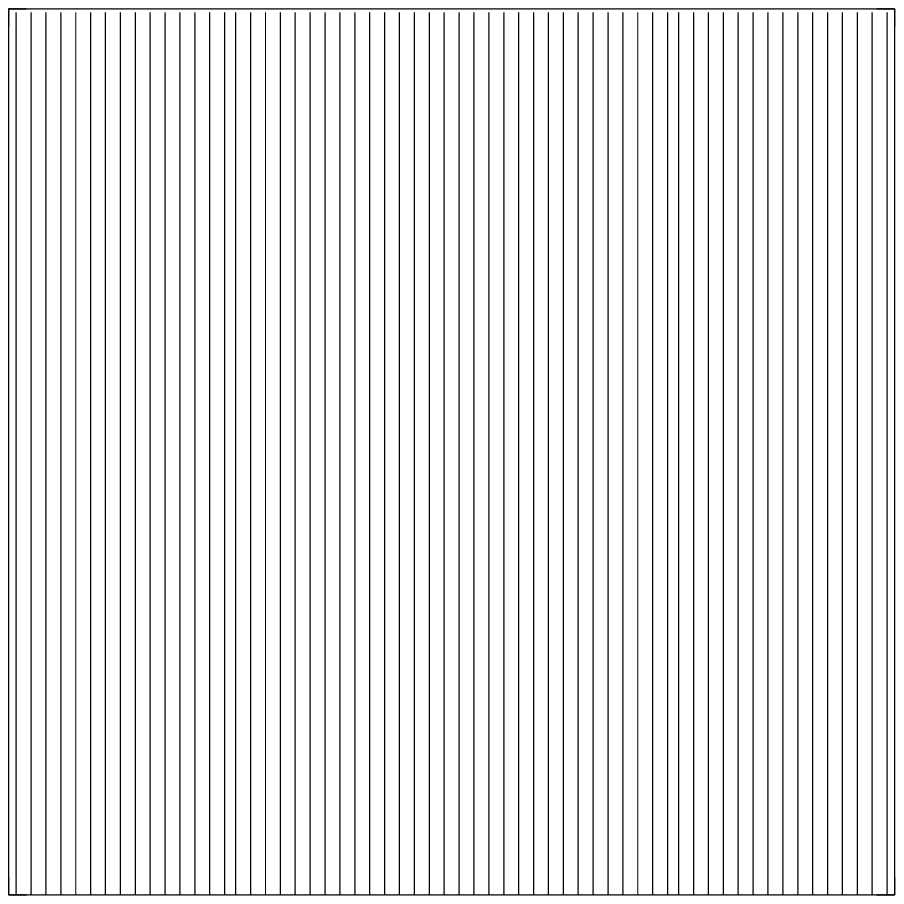}
\includegraphics[scale=0.5]{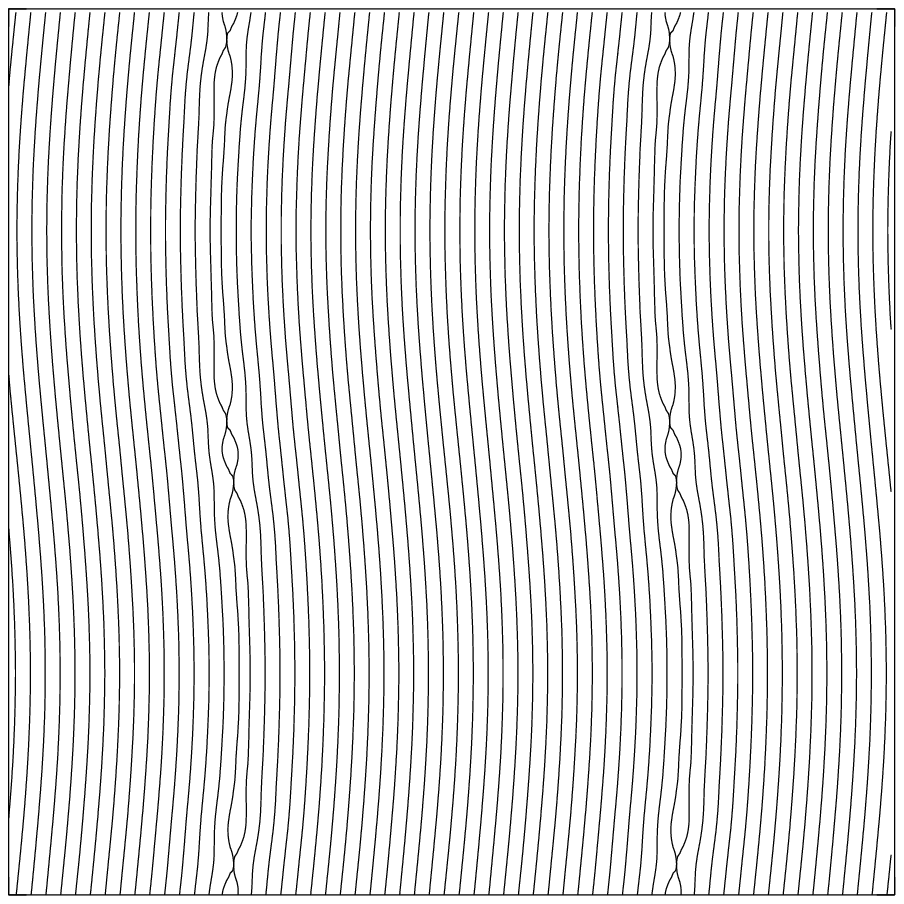}
\includegraphics[scale=0.5]{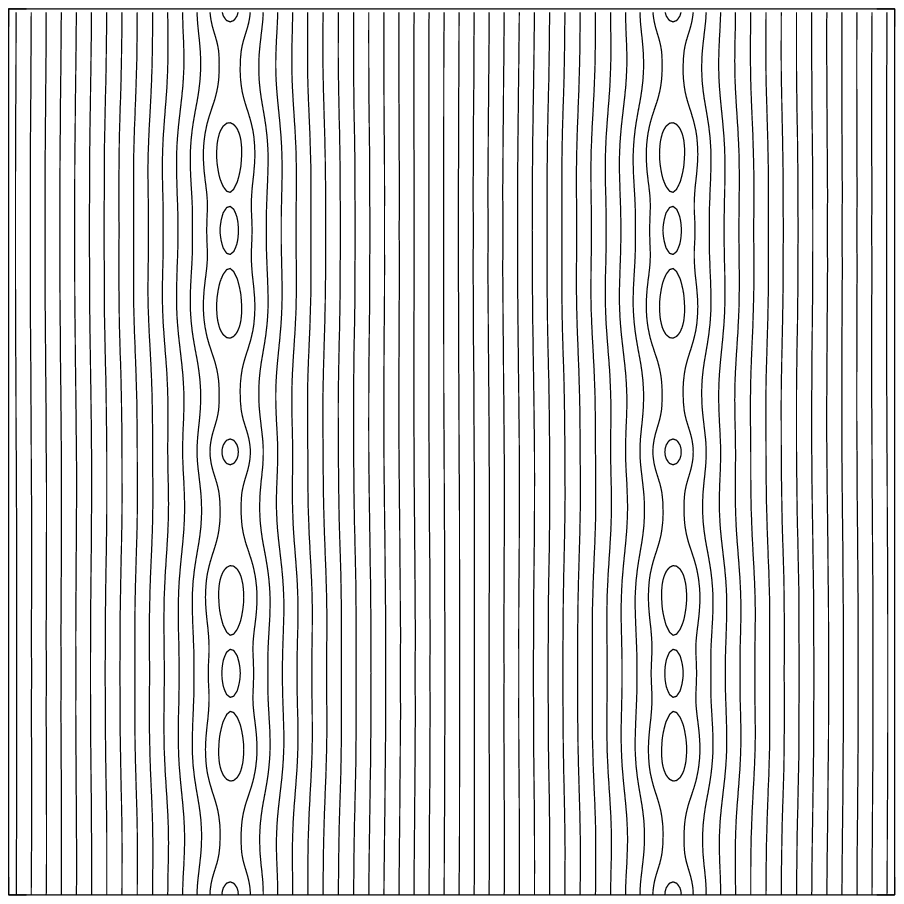}
\includegraphics[scale=0.5]{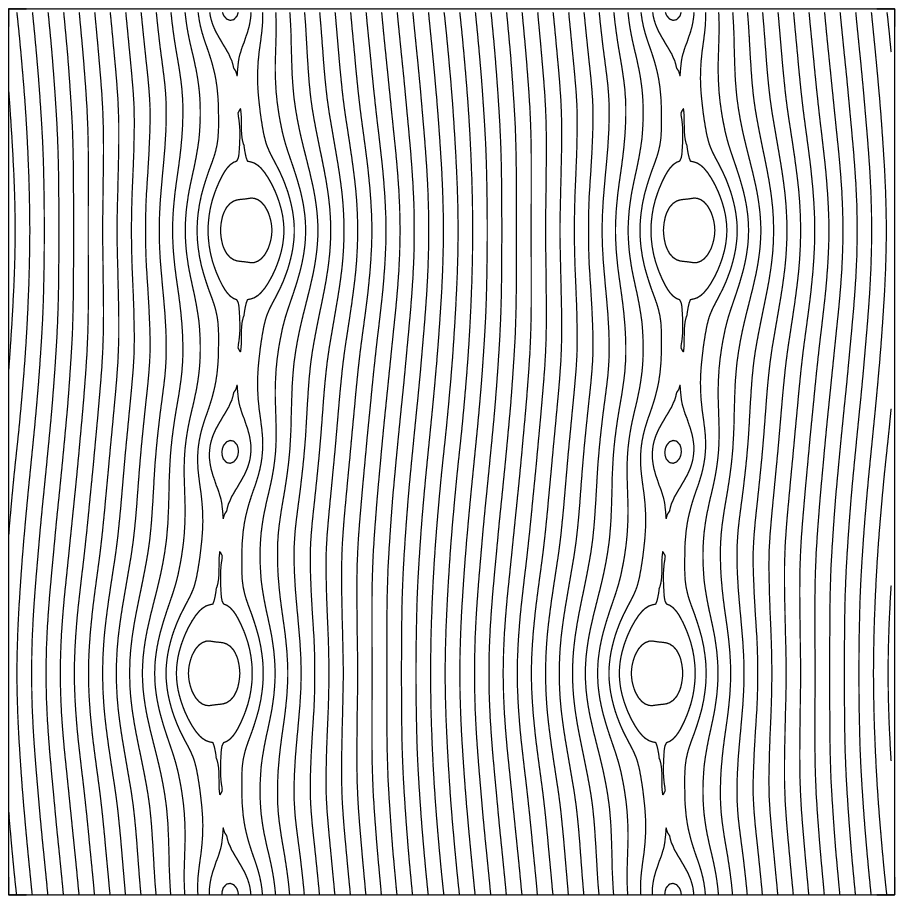}
\includegraphics[scale=0.5]{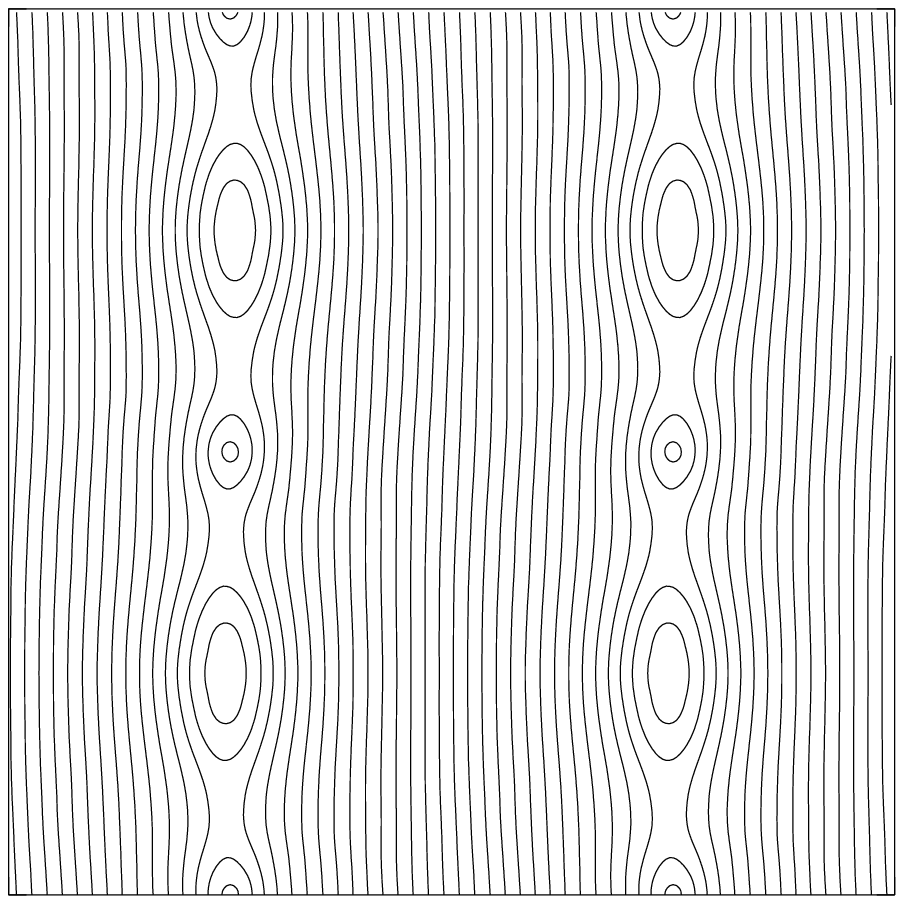}
\includegraphics[scale=0.5]{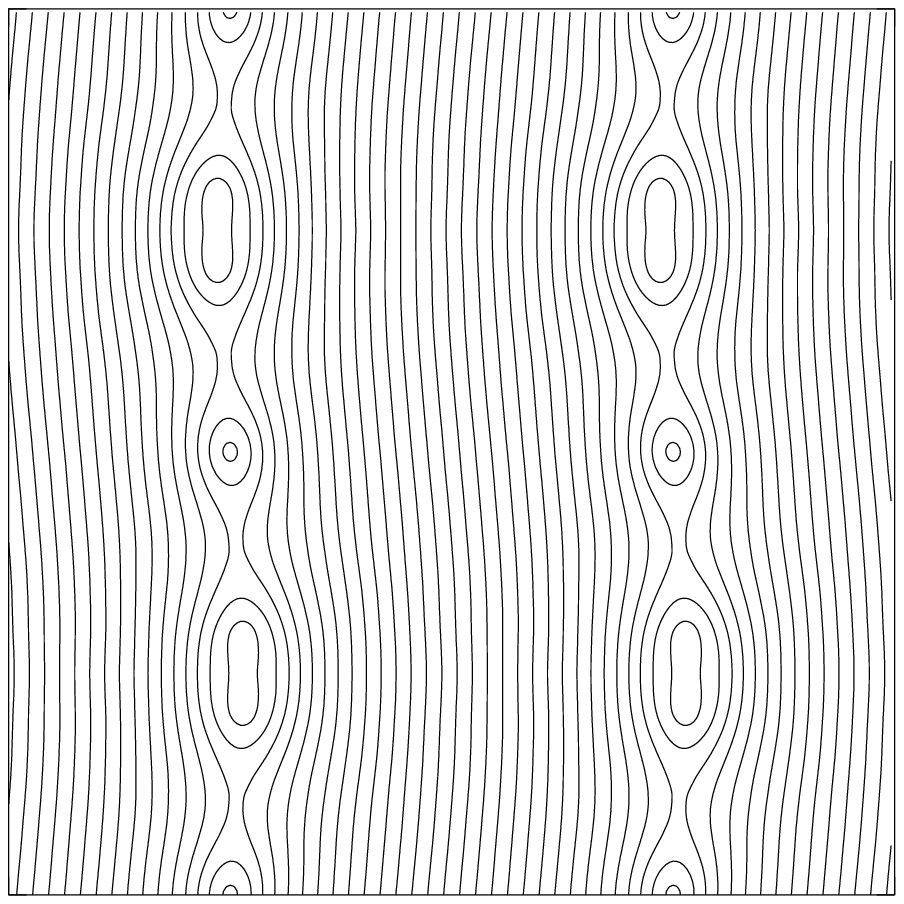}
\includegraphics[scale=0.5]{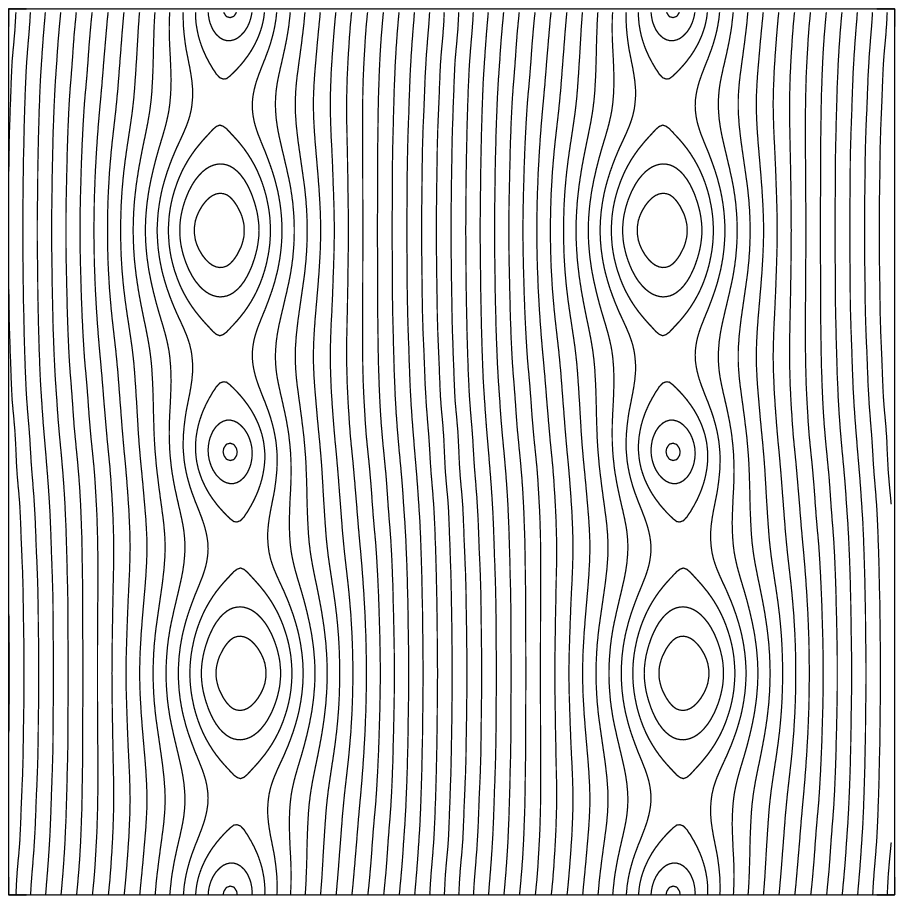}
\includegraphics[scale=0.5]{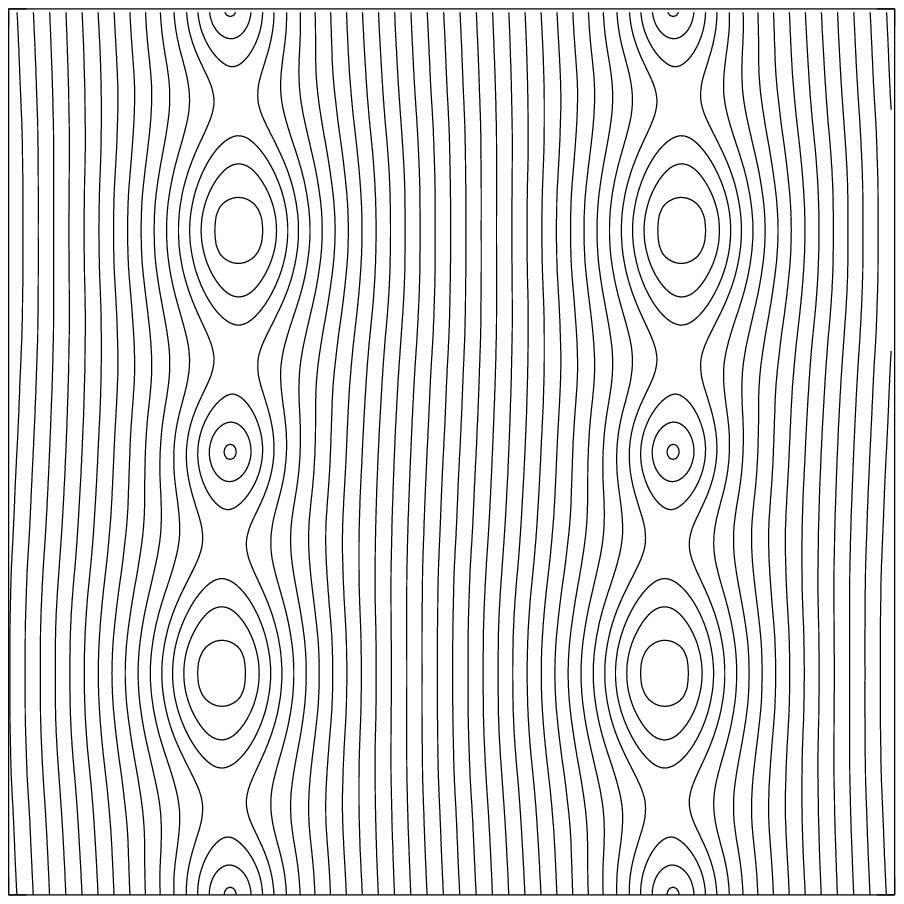}
\includegraphics[scale=0.5]{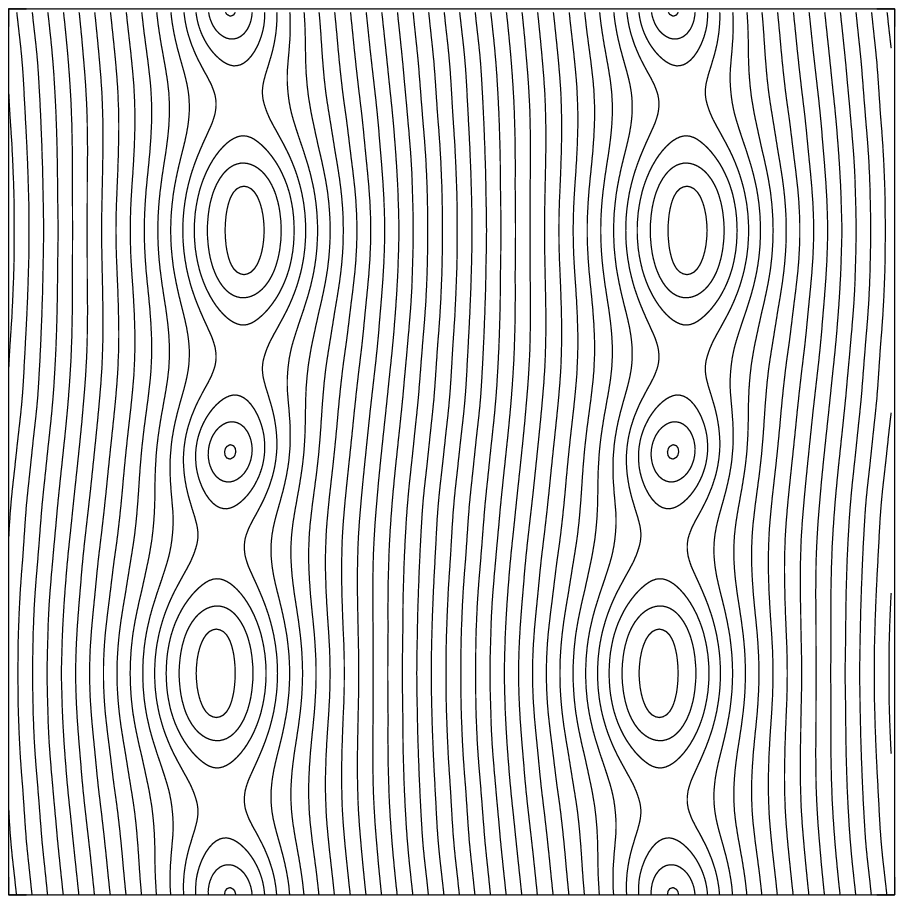}
\caption{Time evolution of the magnetic field lines during the current
sheet test. The  calculation was performed on a  uniform grid composed
of $256^2$ cells,  using the Roe solver and  the MonCen slope limiter.
From top left to  bottom right, the snapshots corresponds successively
to times  $t=0,0.5,1,1.5,2,2.5,3, 3.5$ and $4$. The  entire figure can
be compared to the figure $12$ of \citet{gardiner&stone05a}.}
\label{current sheet}
\end{center}
\end{figure*}

\citet{gardiner&stone05a}  recently described a  2D test  that follows
the  time evolution of  a current  sheet created  by a  magnetic field
discontinuity.    Reconnection   occurs  at   the   location  of   the
discontinuity.   Because no  explicit dissipation  is included  in the
code, the entire evolution is  driven by the numerical resisitivity of
the  scheme,  and, as  such,  is sensitive  to  every  details of  the
algorithm. The initial setup is described in the followings.

\begin{figure*}
\begin{center}
\includegraphics[scale=0.87]{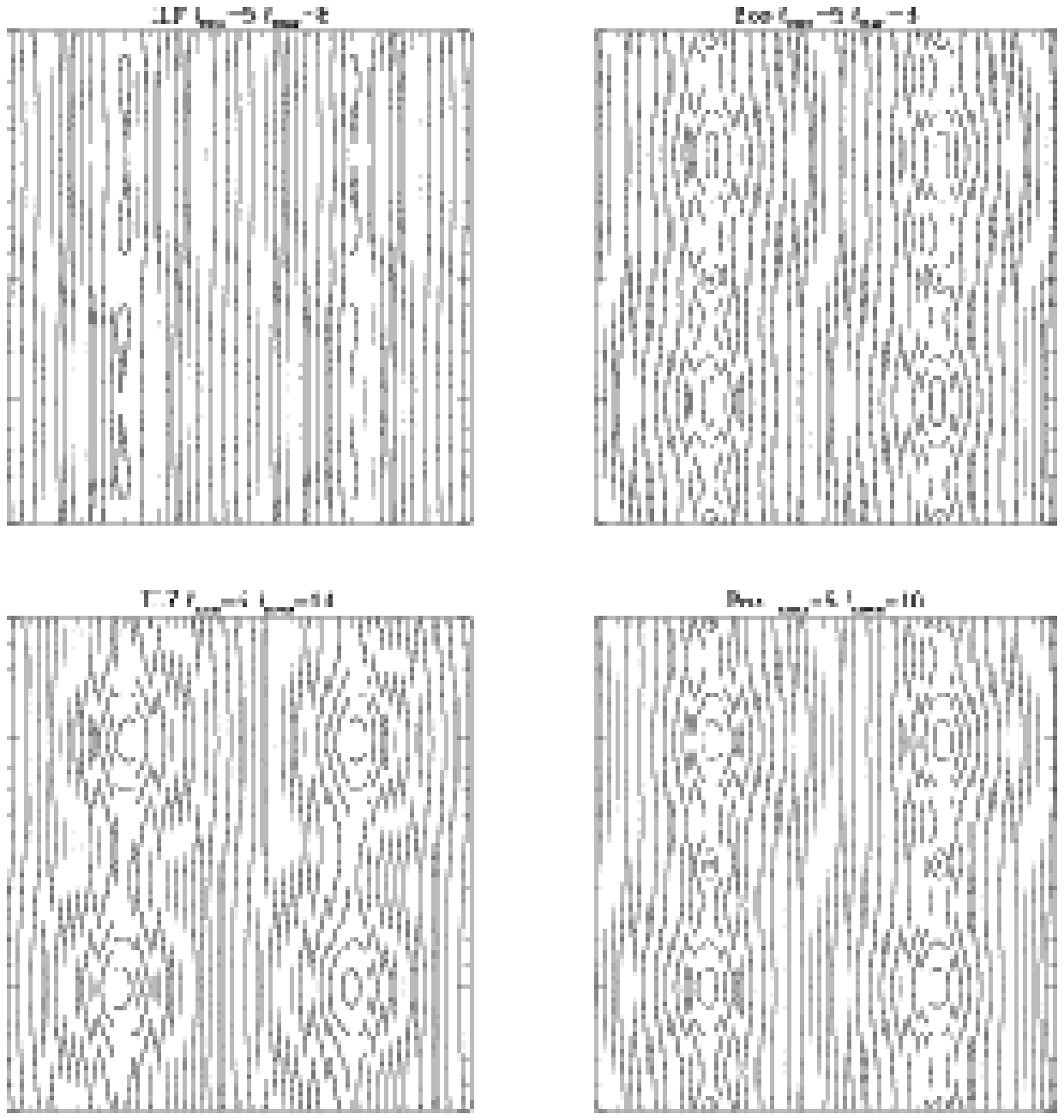}
\caption{This  figure compares  the magnetic  field lines  obtained at
time $t=2$  for high--resolution  runs (bottom) and  low--resolution runs
(top),  as  well  as for  the  Roe  solver  (right) and  the  local
Lax--Friedrich solver (left). Both low and high--resolution runs
  were performed using the AMR scheme: $l_{min}=5$ for the former and
  $l_{min}=10$ for the latter. The refinement strategy is detailed in
  the text.}
\label{sheet convergence}
\end{center}
\end{figure*}

The computational domain  lies in the domain $0 \leq x  \leq 2$ and $0
\leq  y  \leq  2$ and  is  divided  in  $256$  uniform cells  in  each
directions. At time $t=0$,  density and pressure are uniform: $\rho=1$
and $P=0.1$. The  magnetic field components vanish in  the $x$ and $z$
direction and $B_y$ is defined by
\begin{eqnarray}
B_y= \left\{ \begin{array}{ll}
-1 & \textrm{if $|x-1| \leq 0.5$}  \, , \\
+1 & \textrm{otherwise} \, .
\end{array} \right.
\end{eqnarray}
Similarly, $v_y=v_z=0$ and $v_x=v_0 \sin~(\pi y)$.

\begin{figure*}
\begin{center}
\includegraphics[scale=0.55]{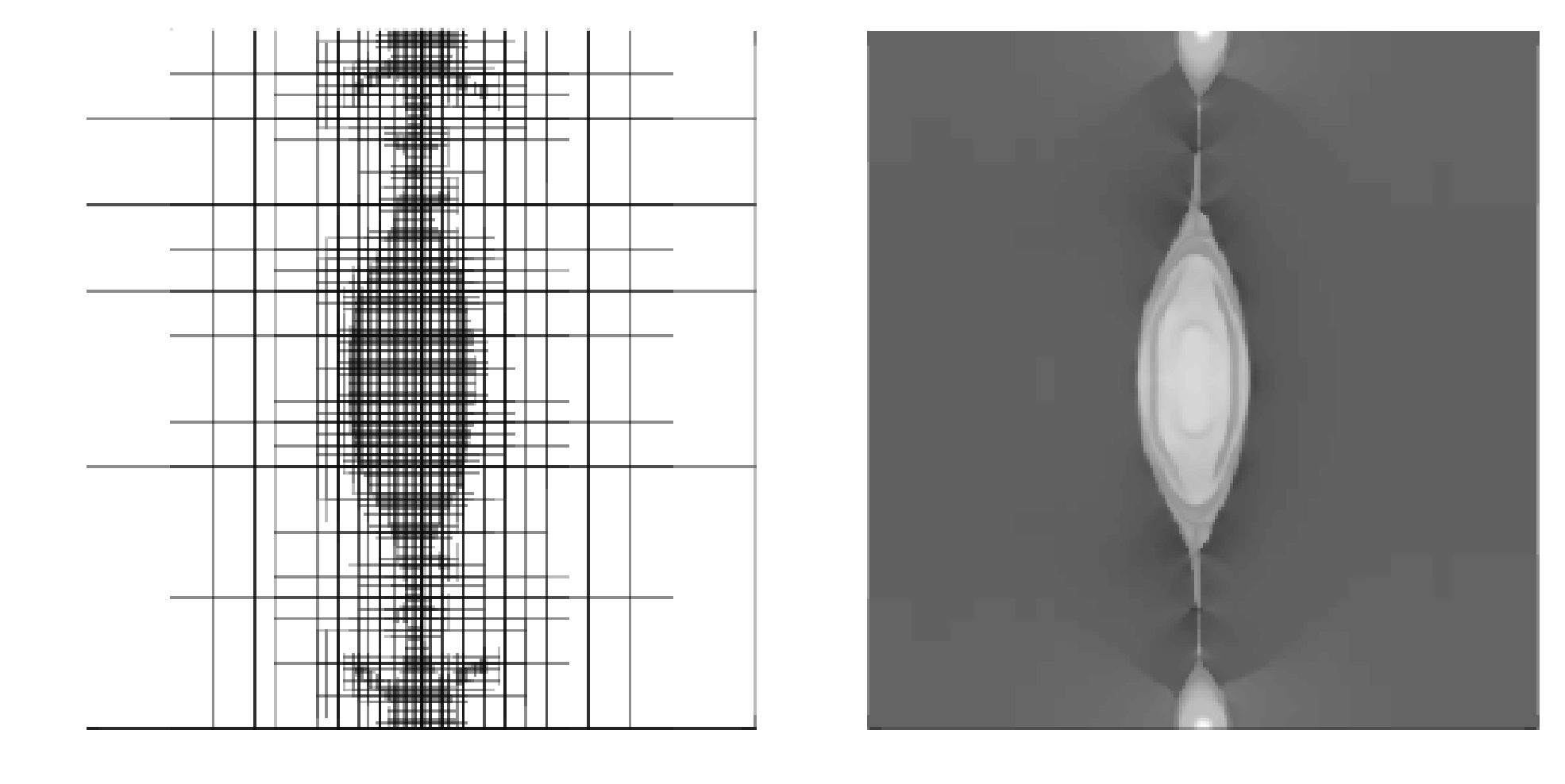}
\caption{This  figure shows  a zoom  on the bottom left magnetic
island for the high--resolution run with the Roe solver at time $t=3$.
On  the  left,  only  octs  boundaries  are  plotted  to  clarify  the
visualization of  the AMR grid.  On  the right, a grey  scale image of
the  thermal pressure  is  shown, with  strong transverse  shock--waves
clearly visible as sharp discontinuities.}
\label{sheet grid}
\end{center}
\end{figure*}

We first present  our numerical results using the  Roe solver with the
MonCen   slope   limiter. They are represented on  figure~\ref{current
  sheet}. The  magnetic field lines
are plotted at times $t=0$,  $0.5$, $1$, $1.5$, $2$, $2.5$, $3$, $3.5$
and $4$.   As reported by  \citet{gardiner&stone05a}, magnetic islands
form, grow  and eventually merge with  each other.  At the  end of the
simulation, four islands  are clearly visible at the  location of each
discontinuity. A direct comparison between our results and figure $12$
of  \citet{gardiner&stone05a}  shows  that  both  codes  agree  almost
perfectly up  to time $t=2.5$.  On  the other hand, at  later time, no
strong evolution is  observed in our case, while  for the ATHENA code,
the flow  symmetry is broken and  the two islands merge  into a single
large  one   \citep{gardiner&stone05a}.   As  discussed   above,  this
difference is  an indication that  both codes, although  very similar,
have different dissipative properties.

The next  step is to test  our AMR scheme  in 2D: we perform  the same
exact simulation, except that now  we use a base grid of $n_x=n_y=32$,
which  corresponds  to  $\ell_{min}=5$,  with  3  additional  levels  of
refinement,  so  that $\ell_{max}=8$.   The  formal  resolution is  thus
equivalent to the  first test with a regular Cartesian  grid. We use a
refinement strategy based on the  gradient of the thermal pressure:
\begin{equation}
\frac{\max \left( \left| \Delta_x P \right|, \left| \Delta_y P \right|
\right) } {P}> 0.05 \, ,
\end{equation}
associated with a similar criterion based on $B_y$.
In Figure~\ref{sheet convergence}, we show the magnetic field lines at
time  $t=2$ obtained with  the Roe  solver (upper  right plot):  it is
indistinguishable from  the previous result obtained  with a Cartesian
grid.   For sake  of  comparison,  we have  also  compared the  result
obtained at $t=2$ with the  Lax-Friedrich solver (upper left plot): it
is  now   completely  different.   Due  to   the  increased  numerical
diffusivity of  the Lax-Friedrich  solver, the tiny  magnetic islands,
that  move up and  down from  the center  of the  image, have  not yet
merged to their final position. Moreover, with the Roe solver, we also
obtain a static magnetic island at the position of the flow stagnation
point. This static island  is absent from the Lax--Friedrich solution.
As anticipated,  this test is  highly discriminant of  the dissipative
properties of numerical schemes. 

In order  to study  the convergence of  each solution, we take
advantage of the
speed--up   provided   by  the   AMR   grid   to  perform   additional
high--resolution simulation.   Using the same  refinement strategy, we
now  set   $\ell_{max}=10$,  so  that  the  formal   resolution  is  now
$n_x=n_y=1024$.  We  present in Figure~\ref{sheet  grid} the resulting
AMR grid, together  with a grey--scale image of  the thermal pressure.
We see that  AMR cells are optimally distributed  in order to properly
sample  the current  sheet,  as  well as  sharp  MHD waves  propagating
perpendicular  to the current  sheet.  The  results obtained  at $t=2$
using  both  Riemann  solvers  are   shown  at  the  bottom  plots  of
Figure~\ref{sheet convergence}.  The Roe solution has not changed when
compared to  its low--resolution counterpart.   This demonstrates that
the  lower numerical  dissipation of  the Roe  solver allows  a faster
convergence   of   the   numerical   solution.    Interestingly,   the
Lax--Friedrich high--resolution solution has also converged toward the
same  solution, except  for the  static  magnetic island  at the  flow
stagnation point.   This demonstrates that using AMR  can provide fast
convergence towards the true  solution, even with a rather dissipative
scheme. On the other hand, the static island in the center of the flow
seems to be highly sensitive to the details of the scheme.  As opposed
to Lax--Friedrich,  the Roe solver  has the interesting  property that
for  a  static  velocity  field, the  numerical  dissipation  vanishes
exactly.  The difference between  the 2 solvers is therefore maximized
at the stagnation  point, where both schemes are  converging towards 2
different  solutions, even  at our  highest resolution.  This peculiar
behavior  is  due  to  the  fact that  this  reconnection  problem  is
performed without  any physical  resistivity.  It should  be therefore
considered only as  an interesting numerical test, rather  than a true
physical application.

\section{Astrophysical applications in 3D}
\label{astro appli}

To illustrate the possibilities of  RAMSES, we present in this section
two  3D  tests  of   astrophysical  significance:  the  development  of
Magneto--Rotational Instability (MRI) and the formation of a magnetized
molecular core. 

\subsection{The magnetorotational instability}

\begin{figure}
\begin{center}
\includegraphics[scale=0.32]{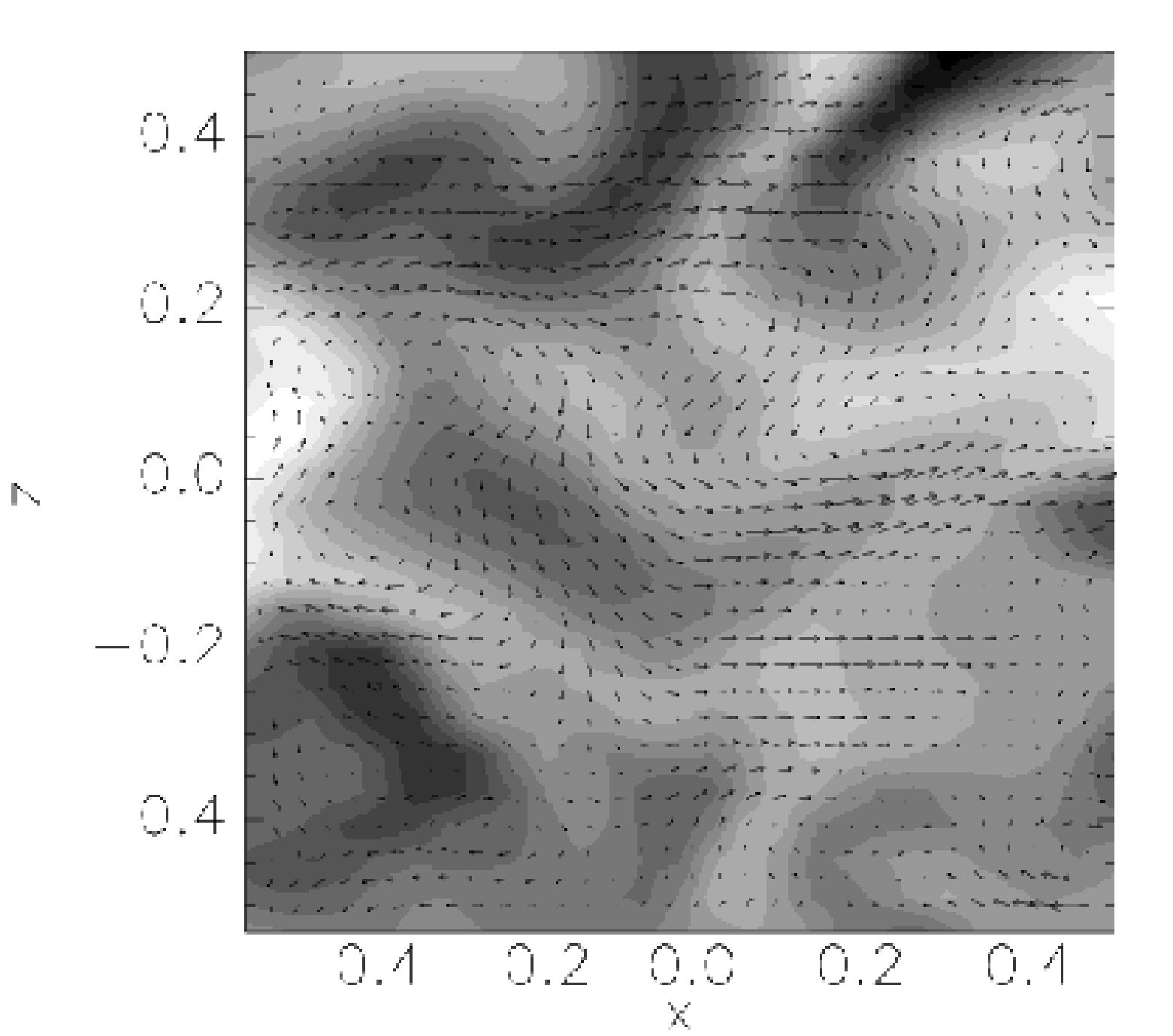}
\caption{Structure of the flow in the $(x,z)$ plane after 60
  orbits. The arrows shows the poloidal velocity field overplotted on
  gray scale contours of the y--component of the magnetic
  field. Because of the growth of the MRI, the entire flow has become
  turbulent.}
\label{mri snapshot}
\end{center}
\end{figure}

\begin{figure}
\begin{center}
\includegraphics[scale=0.5]{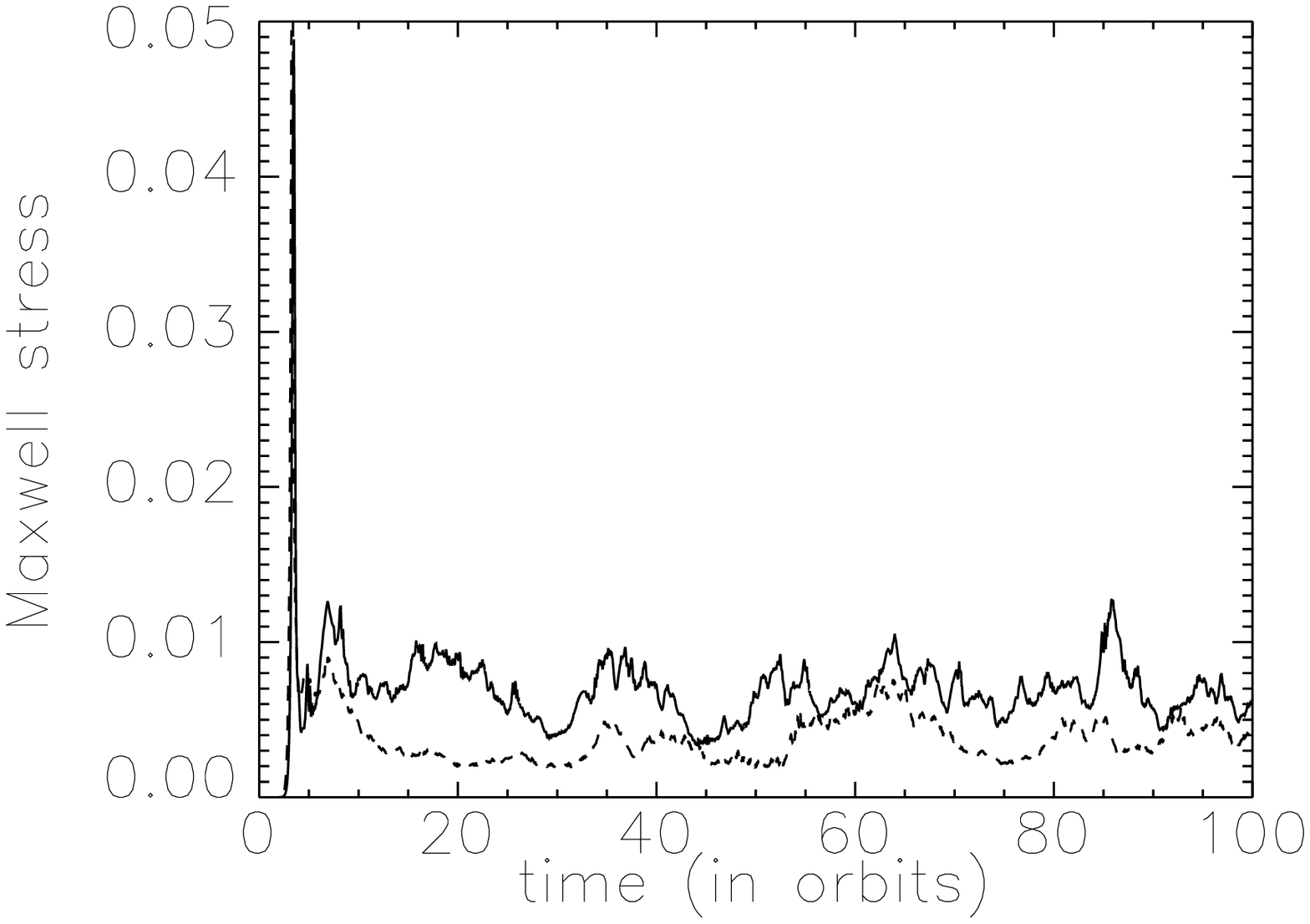}
\caption{Time history of the volume averaged Maxwell stress tensor 
normalized by pressure, obtained in the shearing box model. The 
solid line was computed using RAMSES and the dashed line was calculated 
using ZEUS. Both models use exactly the same set of parameters. Both 
shows sustained MHD turbulence and a similar amount of angular
momentum transport.}
\label{mri time history}
\end{center}
\end{figure}

The development of MHD turbulence resulting from the growth of the MRI
is likely  to be at  the origin of an efficient radial  transport in
accretion disks.  This instability  has been extensively studied using
finite difference  codes like  ZEUS \citep{hawleyetal95}; our  goal is
here to compare  the results of RAMSES with  those obtained in earlier
studies obtained in the last decade.

The   MRI   is  a   linear   instability   that   was  first
discovered in the 60s \citep{velikhov59,chandrabook61} before being
applied to accretion disks theory by \citet{balbus&hawley91}. It
operates in rotating flows threaded by a
weak  magnetic  field when  the  angular  velocity decreases  outward.
Numerical  simulations applied to accretion disks have  shown  that
the linear  growth  of  the
instability  is followed  by  MHD turbulence  that transports  angular
momentum outward in the disk, thereby solving a long standing issue in
accretion     disk    theory     (see    \citeauthor{balbus&hawley98},
\citeyear{balbus&hawley98}  for  a  review).  One  subclass  of  these
simulations  has  been  realised  using the  so--called  shearing  box
approximation.  It is a local expansion of the dynamical equation in a
Cartesian  box  around  a  particular  radius of  the  accretion  disk
\citep{goldreich&lyndenbell65}. The interest of this local approach is
that it  is able  to capture  the dynamics of  the accretion  disk and
enables large resolution to be achieved  at the same time. With such a
model,  the  properties of  the  MHD  turbulence  can be  rather  well
studied.

So  far, most of  the shearing  box simulations  have been  done using
artificial viscosity codes,  like  ZEUS  \citep{hawleyetal95},  NIRVANA
\citep{papaloizouetal04} or the Pencil Code \citep{brandenburgetal95}.
Recently, \citet{gardiner&stone05b} applied the ATHENA code to
the  same  problem. They found that using a Riemann solver make little
difference.

We have  implemented the shearing box  model in RAMSES. To  do so, two
main  modifications  have  to  be  made.  First,  the  inertial  terms
appearing in  the equations are  treated as source terms.  Second, the
boundary  conditions  need to  be  adapted  to  the model.   They  are
periodic in $y$  and $z$, which respectively correspond  to the $\phi$
and  $z$  coordinates of  the  standard  cylindrical coordinates.  The
boundary  conditions in  $x$  (the equivalent  of  $R$ in  cylindrical
coordinates)  are the so--called  shearing boundary  conditions widely
discussed in the literature \citep{hawleyetal95,gardiner&stone05b}.

The initial conditions of our  runs are those of the standard shearing
box  model.  The  initial  density  is  uniform  and  equal  to  unity
everywhere. The velocity is such that
\begin{eqnarray}
\bb{v}=\left( 
\begin{array}{c}
0 \\
-q \Omega_0 x \\
0
\end{array} \, ,
\right)
\end{eqnarray}
with $q=1.5$ and $\Omega_0=10^{-3}$. The equation of state is isothermal:
$P=\rho c_{0}^2$, with $c_{0}=10^{-3}$. The initial magnetic field is 
initially purely vertical and its intensity varies sinusoidally with $x$ 
such that the total net flux threading the computational domain vanishes:
\begin{equation}
B_z=B_0 \sin 2 \pi x \, .
\end{equation}
$B_0$ is  calculated such  that the ratio  $\beta$ between  the volume
averaged thermal  to magnetic pressure equals $400$.  The uniform grid
satisfies $-0.5  \leq x \leq  0.5$, $0 \leq  y \leq 2\pi y$  and $-0.5
\leq z  \leq 0.5$.  The resolution is  $(N_x,N_y,N_z)=(32,100,32)$. We
ran the  model with RAMSES, using  the Roe Riemann  solver and the
MonCen slope limiter.

At $t=0$, small random velocities are superposed on the initial state.
The  model  is  evolved  during  $100$ orbital  periods  $T_0$,  where
$T_0=2\pi/\Omega_0$. During the first five orbits, the magnetic energy
is observed to grow. A measure of the growth rate of the
  instability $\sigma$ during that period gives $\sigma \sim 0.55
  \Omega$. It is difficult to compare this growth rate with the
  results of linear theory: since the vertical magnetic field varies with
$x$, its growth rate does not correspond to that of a single
  normal mode. Nevertheless, we expect the volume averaged evolution
  of the magnetic field to be dominated by the growth of the field at
  the position where its strength is initially the largest. This is 
  confirmed by a visual inspection of the structure of the flow
  during that phase, which also indicates that the associated
  wavelengths in the radial and vertical direction are respectively
  equal to $H$ and $H/2$. The results of linear theory
  \citep{balbus&hawley91} predicts that $\sigma_{th}=0.55 \Omega$ in
  that case. Although there is good agreement between $\sigma$ and
  $\sigma_{th}$, we want to emphasize that the treatment presented here is
  only approximate. It nevertheless gives confidence in the results of
  the numerical simulation.

When nonlinear effect becomes important, the magnetic energy reaches
  a peak and start to decline as 
the  whole flow  breaks down  into turbulence,  before levelling  to a
quasi  steady state  it keeps  until the  end of  the  simulation. The
turbulent  nature  of  the  disk  is  illustrated  on  figure~\ref{mri
snapshot}. It  shows the  structure of the  flow in the  $(x,z)$ plane
after $60$  orbits. The arrows  represent the poloidal  velocity field
and  are overplotted on  gray scale  contours of  $B_y$, which  is the
dominant component of the magnetic  field. In order to better quantify
the  strength  of the  turbulence,  we  plot  on figure~\ref{mri  time
history}  the  time history  of  the  volume  averaged Maxwell  stress
tensor, normalised by the mean thermal pressure $P_0$:
\begin{equation}
T_{r\phi}^{Max}=-\frac{<B_x B_y >}{P_0} \, ,
\end{equation}
where $<.>$  denotes a volume average.  The solid  line in
figure~\ref{mri time history} was  obtained with RAMSES.  As a
comparison, the  dashed line
shows the same quantity obtained with ZEUS for the same model 
  (we note that the growth rate measured during the linear phase in
  that case was
  found to be identical to that obtained with RAMSES). The two
curves are in good agreement  with each other, even if there is a
tendency  for RAMSES  to  display some  more  activity.  Indeed,  time
averaging  the  curves  presented  on  figure~\ref{mri  time  history}
between  $40$  and $100$  orbits,  we  obtained  mean values  and  rms
deviation for the Maxwell stress tensor equal to $(6.6 \pm 1.5) \times
10^{-3}$ with RAMSES and equal  to $(3.9 \pm 1.3) \times 10^{-3}$ with
ZEUS. It is worth noting that the (small) difference between
  these two values may not be significant as it is an effect of the
  different dissipative
  properties of the codes: turbulence is driven on large scales by the
  MRI and damps on small scales due to numerical dissipation. The
  precise saturated value of the Maxwell stress results from a 
  balance between these two.

\subsection{Magnetized cloud core collapse}
Here we present another 3D test of astrophysical significance: the 
magnetized collapse of a dense prestellar core. In this problem the  
AMR  scheme is very useful since the density varies over 8 orders of
magnitude and the spacial scale, which is about the Jeans length,
varies over 3-4 orders of magnitude.

Such calculations in the hydrodynamical case have been performed by
several authors using either SPH methods (e.g. \citeauthor{hennebelle04} 
\citeyear{hennebelle04}) or grid
based method (e.g. \citeauthor{matsumoto&hanawa03}
\citeyear{matsumoto&hanawa03}). In the magnetized case, much fewer 3D
calculations have been carried out using SPH
\citep{hosking&whitworth04}, nested grid (Machida et al. 2005a,b, 
\citeauthor{banarjeepudritz06} \citeyear{banarjeepudritz06}) or an AMR
implementation \citep{ziegler05}.

In order to do a precise comparison, we adopt the same initial conditions 
as Hosking \& Whitworth (2005) and Ziegler (2005). 
The cloud has initially a uniform density of $\rho = 4.8 \times 10^{-18}$
g cm$^{-3}$, a temperature of 10 K and a radius of $R_c=0.015$ pc.
The total mass is equal to one solar mass and the initial ratio 
of thermal to gravitational energy is about $0.35$. 
The cloud is initially in solid body rotation with angular 
velocity $\omega = 4.25 \times 10^{-13}$ s$^{-1}$ leading to an 
initial ratio of rotational to gravitational energy of about 0.45.
To induce fragmentation, an $m=2$ perturbation on the density field
 with an amplitude of 10$\%$ has been setup initially.
The magnetic field is initially uniform and parallel to the rotation axis.
%The initial mass-to-flux over critical mass-to-flux ratio of the core is: 
%$\mu=\sqrt{G} M / \pi R^2 B_{z,0}$. 
We use the same barotropic equation of state as Hosking \& Whitworth (2005), 
namely $P = C_s^2 \rho \times (1 + (\rho / \rho_{\rm crit})^{4/3})^{1/2} $ with
$\rho_{\rm crit}=10^{-13}$ g cm$^{-3}$. All calculations have
initially $64^3$ cells and uses 8 additional AMR levels. The
refinement criteria
is based on the Jeans length which  is numerically described by at
least 10 cells.

\subsubsection{Hydrodynamical collapse}

\begin{figure}
\begin{center}
\includegraphics[scale=0.95]{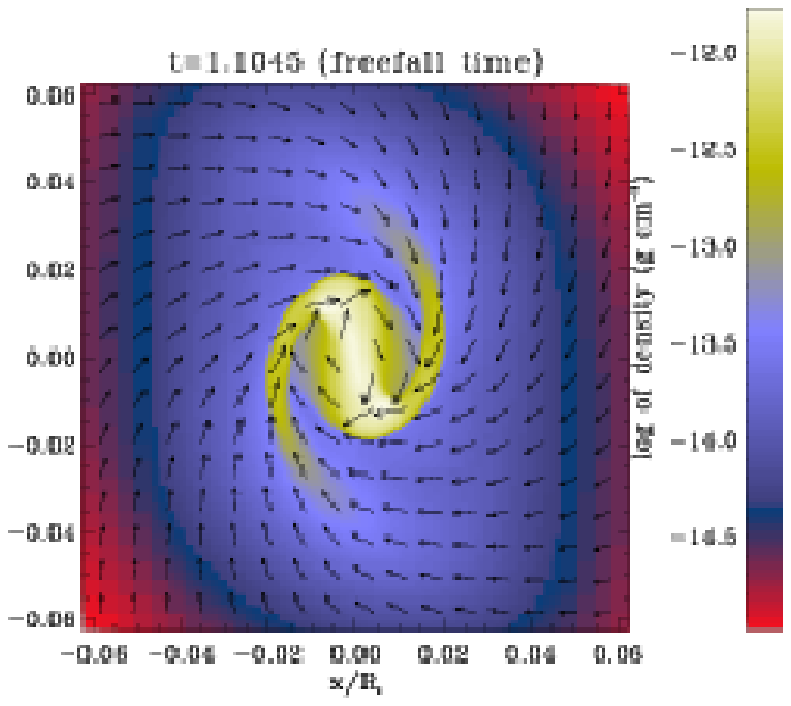}
\includegraphics[scale=0.95]{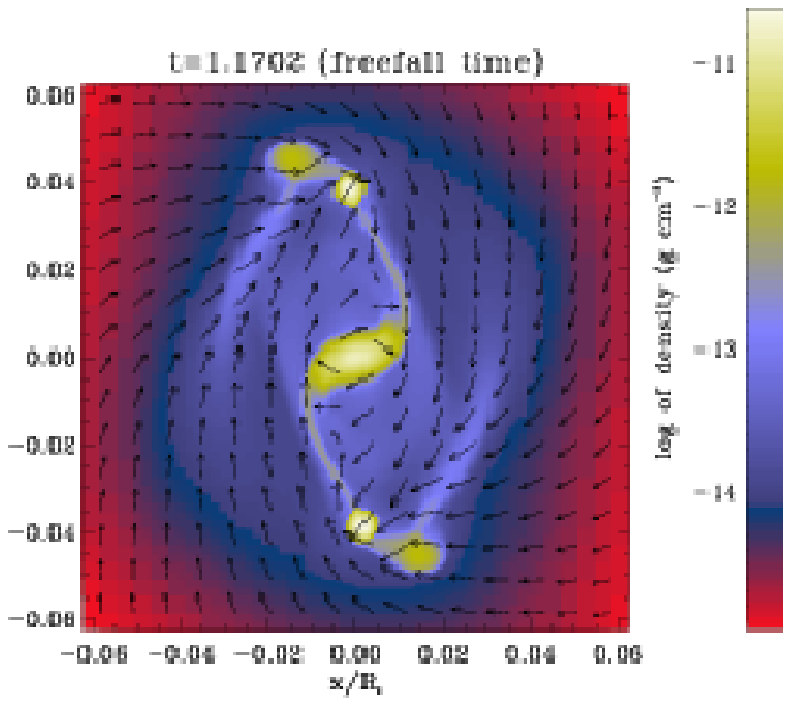}
\includegraphics[scale=0.95]{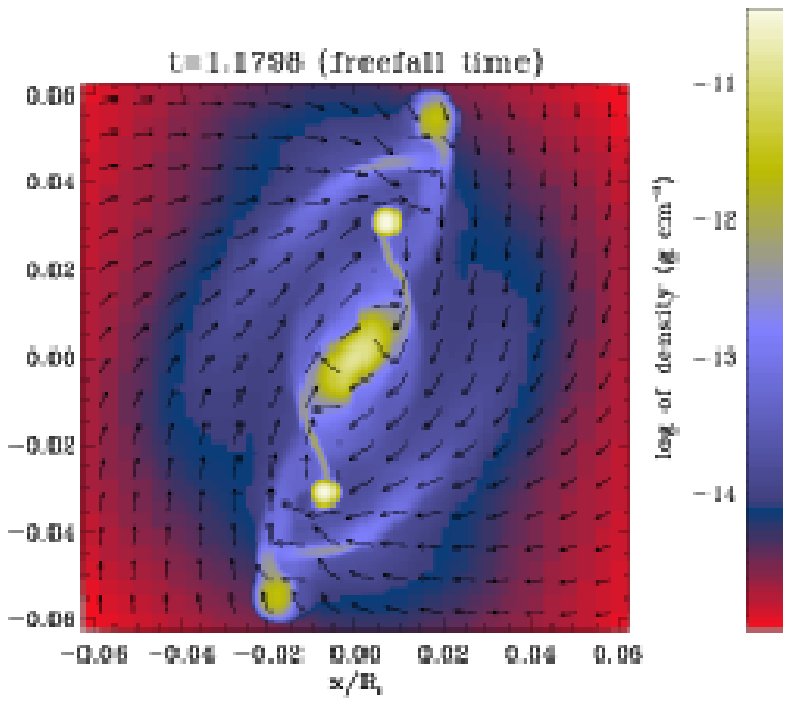}
\end{center}
\caption{Three timesteps illustrating the hydrodynamical collapse.}
\label{hydrocol}
\end{figure}

Figure~\ref{hydrocol} displays  three snapshots of the 
hydrodynamical case performed with the Lax-Friedrich solver. The equatorial
density field is displayed and the arrows show the velocity field.
The top panel is very similar to the second panel of Fig.~8 of Ziegler (2005).
Both spiral patterns have approximately the same size and the same 
orientation.  
Timing is also very similar (agreement within about 5$\%$ of accuracy).
The result compares well to the SPH calculation of Hosking \& Whitworth (2005) shown in their Fig.~2 (bottom-right panel).
 The second panel of Fig.~\ref{hydrocol} appears to be similar to 
the bottom-left panel of Fig.~3 of Hosking \& Whitworth (2005). 
In both cases about 5 fragments have formed located approximately at the same 
position in both simulations. The third panel shows the density field 
0.01 Myr after the time of the second panel which is comparable to the 
timeshift between left-bottom and right-bottom of Fig.~3 of 
Hosking \& Whitworth (2005). The agreement is less good than 
for earlier time. Our results remain symmetric 
which is not the case for the Hosking \& Whitworth's results. 

\subsubsection{MHD collapse}

In this section we present results for the MHD collapse. 
The intensity of the magnetic field is such that the cloud mass-to-flux
ratio is twice the critical mass-to-flux ratio. This corresponds to the first 
MHD case  studied by Ziegler (2005). 

\begin{figure*}
\begin{center}
 \includegraphics[scale=0.96]{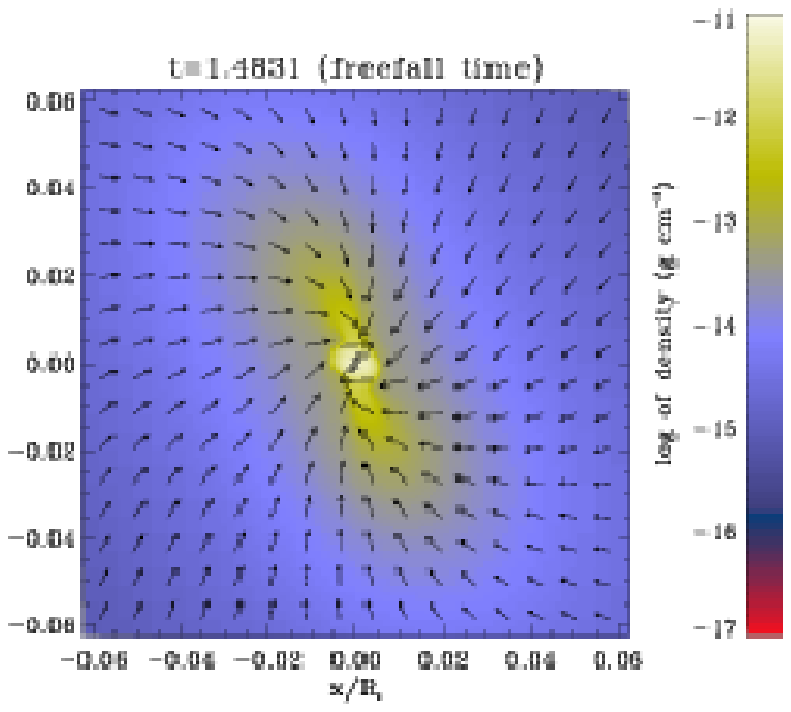}
\hspace{0.4cm}
\includegraphics[scale=0.96]{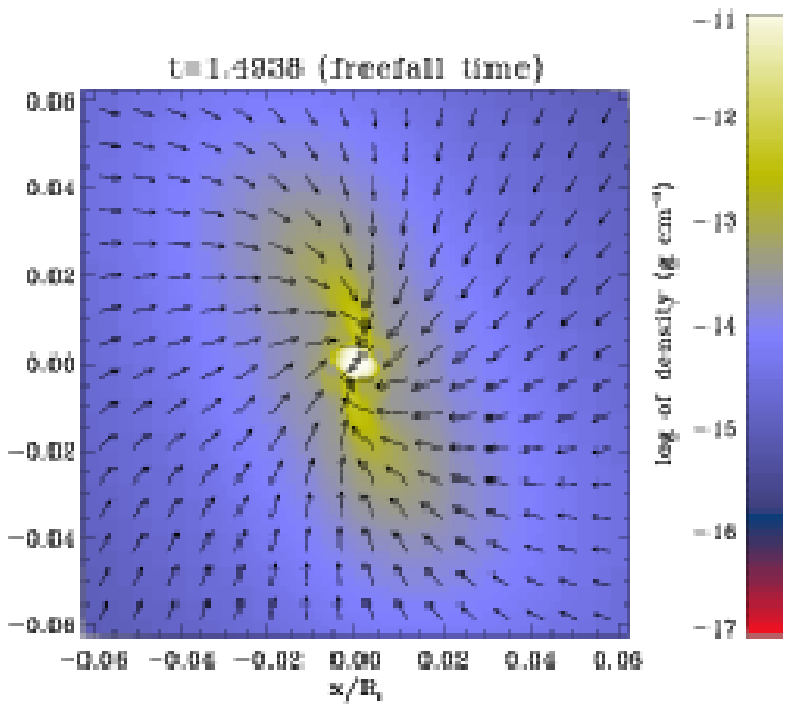}
 \includegraphics[scale=0.96]{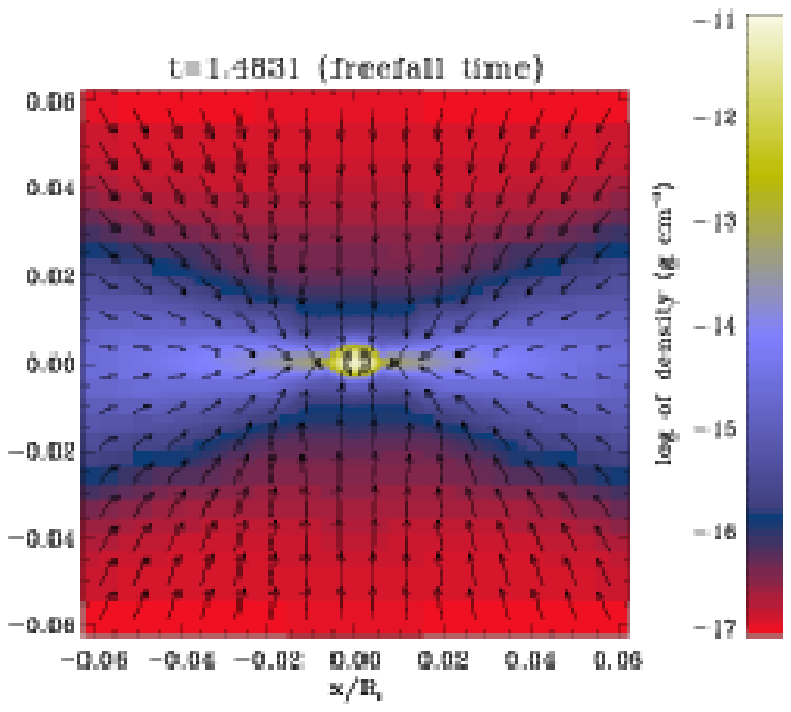}
\hspace{0.4cm}
\includegraphics[scale=0.96]{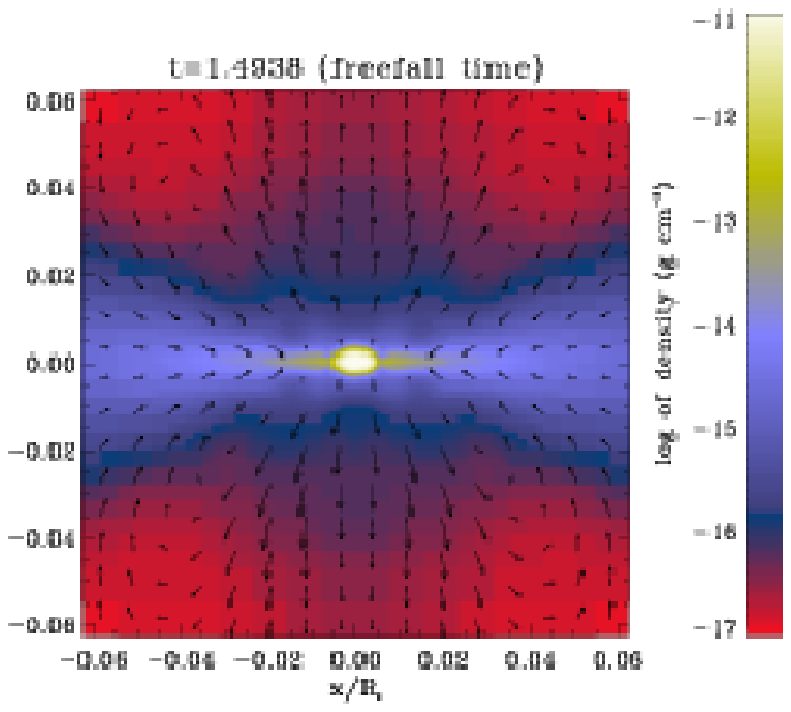}
\caption{Two timesteps illustrating the magnetized collapse. 
The upper panels display the equatorial density and velocity field
whereas bottom panels displays the density in $x-z$ plan. The calculation
is performed with the Lax-Friedrich solver.}
\label{lf_mhd}
\end{center}
\end{figure*}

\begin{figure*}
\begin{center}
\includegraphics[scale=0.95]{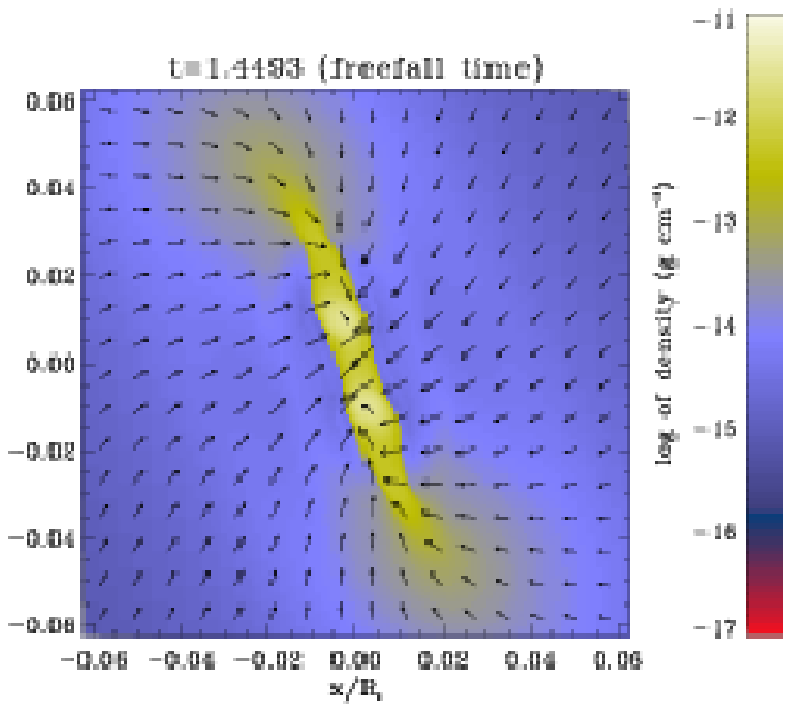}
\hspace{0.4cm}
\includegraphics[scale=0.95]{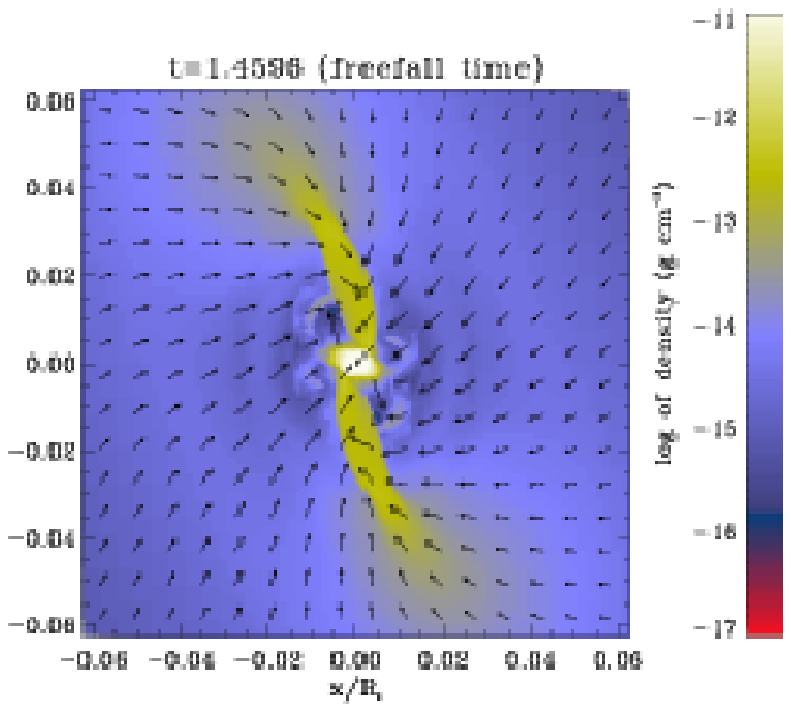}
\includegraphics[scale=0.95]{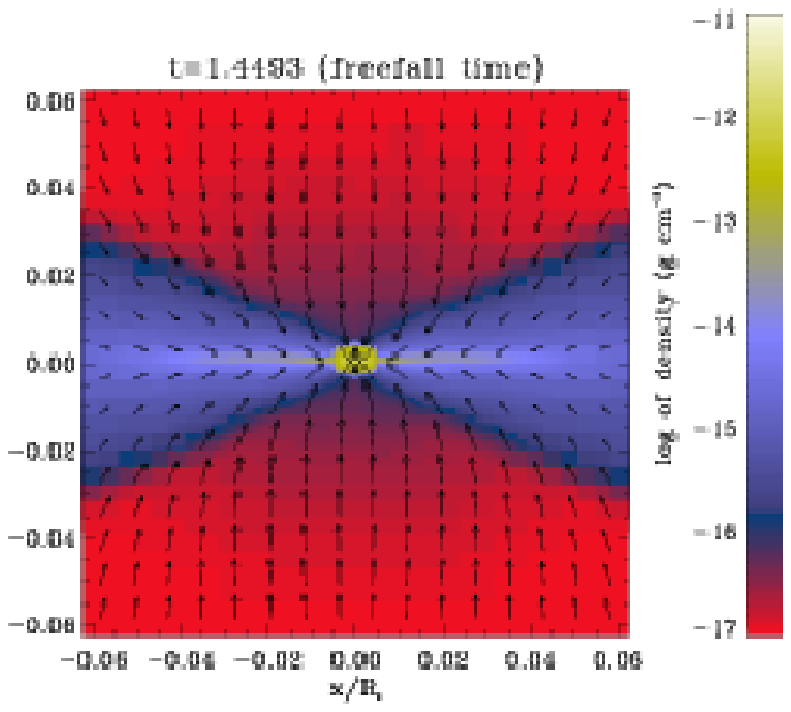}
\hspace{0.4cm}
\includegraphics[scale=0.95]{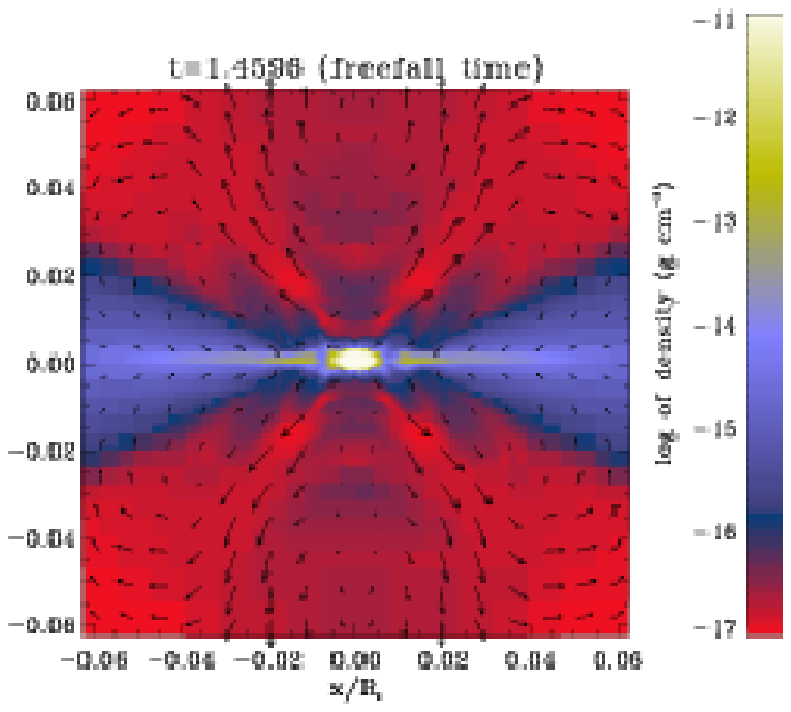}
\caption{Same as Fig.~\ref{lf_mhd} except that the calculation has been  carried out with the Roe solver.}
\label{roe_mhd}
\end{center}
\end{figure*}

Figure~\ref{lf_mhd} shows results for a calculation performed with the 
Lax-Friedrich solver whereas Fig.~\ref{roe_mhd} shows results obtained
with the Roe solver. Upper panels of Fig.~\ref{lf_mhd} and~\ref{roe_mhd}
display the equatorial density 
field whereas bottom panels display the density in the x-z plane.
Left panels display results at a time close to the first panel of
Fig.~10 of Ziegler (2005). 
Right panels display results about 0.01 Myr later. 

In his simulation, Ziegler reports the formation of a binary having a 
separation at time $t \simeq 1.44 t_{\rm ff}$ of about 0.06 $R_c$ where 
$t_{ff}$ is the freefall time.   
As shown in Fig.~\ref{lf_mhd}, there are no binaries with the Lax-Friedrich 
solver. There is also a shift in time (of about $3\%$) since the collapse 
occurs slightly later than $1.44 t_{ff}$. On the contrary, as shown 
in Fig.~\ref{roe_mhd}, in the simulation performed with the Roe solver, 
a binary forms at time $t=1.45 t_{ff}$ although we find a separation 
of 0.03 $R_c$ instead of 0.06 $R_c$. Ziegler reports also another case with 
stronger magnetic field for which the mass-to-flux ratio is 1.2 times 
the critical mass-to-flux ratio. In this case he finds no binary. We have 
also performed this simulation (not displayed here for conciseness) 
and we confirm the absence of binary. Although the formation of the binary 
appears to be a good numerical test, it should be said that it is somehow 
physically artificial since it relies on  initial conditions for which 
the density field is perturbed but not the magnetic field. As a
result, the magnetic Jeans mass at the density maximum is lowered,
thus making the perturbation very unstable. Indeed we have performed a
simulation in which the $m=2$ perturbation has been applied to the
magnetic field as well.  In this case we find no binary.

Right panels of Fig.~\ref{roe_mhd} show that the two 
fragments have merged leaving a single central fragment. This is an
important difference with the hydrodynamical case in which five fragments 
have been found (although further evolution reveals that two of these fragments
 merge with the central one). Another important departure from the 
hydrodynamical case is the presence of strong outflows (right-bottom panel).
There are very similar to the outflows shown in Machida et al. (2005b).
Outflows are also obtained with the Lax-friedrich solver 
(right-botton panel of Fig.~\ref{lf_mhd}) although the flow structure 
is slightly different. With the Roe solver, the velocity 
field along the z-axis vanishes whereas this is not the case with the 
Lax-Friedrich solver. The disk appears to be thicker with 
the Lax-Friedrich solver than with the Roe solver.

We conclude that RAMSES-MHD is able to reproduce quantitatively 
results obtained by various authors including fragmentation and 
outflows. Significant differences appear between results 
obtained with the  Lax-Friedrich and the Roe solvers although the former
is able to reproduce the main features of the flow.  

\section{Conclusion and Perspectives}
\label{conclusion}

In this paper, we have presented an extension of RAMSES to MHD. The
algorithm is based on the MUSCL-Hancock approach already used in the
hydrodynamic version of RAMSES \citep{teyssier02}. The induction
equation is evolved in time using the standard CT scheme
\citep{evans&hawley88}. To do so, time averaged EMFs are computed on
cell edges by solving a 2D Riemann problem, as described in
\citet{londrillo&delzanna00}. Several tests are presented that
illustrate the properties and robustness of the code. In particular,
we show that the AMR scheme implemented in RAMSES can be crucial to
describe accurately the propagation of some unusual waves peculiar to
MHD like the compound waves. 

We also demonstrate the versatility of RAMSES
by studying two problems of astrophysical significance: the
development of MHD turbulence in accretion disk and the collapse of
dense core in the interstellar medium. In both cases, we report
results that are consistent with previous studies published in the
literature. These two applications show that RAMSES is well suited to
study a wide variety of problems involving MHD in astrophysics.

In future studies, several improvements will now be investigated. It
will be particularly useful, for example, to develop a proper 2D
Riemann solver to
calculate the time averaged EMFs, instead of making linear combination
of 1D solvers as it is done now. Nonlinear Riemann solvers could also
be implemented, like HLLC \citep{miyoshi&kusano05} for example. Obviously,
an extension to curvilinear coordinates would also be very interesting,
particularly for applications involving accretion disks or galaxies.
Finally, it will be necessary in some cases to go beyond the
ideal MHD framework and to implement new physics like ohmic
dissipation or ambipolar diffusion.

\section*{ACKNOWLEDGMENTS}
The authors thank Emmanuel Dormy and St\'ephane Colombi for useful
discussions. Some of the simulations presented in this paper were
performed on the QMUL High Performance Computing Facility purchased
under the SRIF initiative and at CCRT, the CEA supercomputing center.

\bibliographystyle{aa}
\bibliography{author}

\newcommand{\noopsort}[1]{}
\begin{thebibliography}{48}
\expandafter\ifx\csname natexlab\endcsname\relax\def\natexlab#1{#1}\fi

\bibitem[{Balbus \& Hawley(1991)}]{balbus&hawley91}
Balbus, S. \& Hawley, J. 1991, ApJ, 376, 214

\bibitem[{Balbus \& Hawley(1998)}]{balbus&hawley98}
Balbus, S. \& Hawley, J. 1998, Rev.Mod.Phys., 70, 1

\bibitem[{Balsara(2001)}]{balsara01}
Balsara, D.~S. 2001, J. Comput. Phys., 174, 614

\bibitem[{{Balsara} \& {Spicer}(1999)}]{balsara&spicer99}
{Balsara}, D.~S. \& {Spicer}, D.~S. 1999, Journal of Computational Physics,
  153, 671

\bibitem[{{Banerjee} \& {Pudritz}(2005)}]{banarjeepudritz06}
{Banerjee}, R. \& {Pudritz}, R. 2005, \apj

\bibitem[{{Berger} \& {Colella}(1989)}]{berger89}
{Berger}, M.~J. \& {Colella}, P. 1989, Journal of Computational Physics, 82, 64

\bibitem[{{Berger} \& {Oliger}(1984)}]{berger84}
{Berger}, M.~J. \& {Oliger}, J. 1984, J. Comp. Phys., 53, 484

\bibitem[{{Brackbill} \& {Barnes}(1980)}]{Brackbill80}
{Brackbill}, J.~U. \& {Barnes}, D.~C. 1980, Journal of Computational Physics,
  35, 426

\bibitem[{{Brandenburg} \& {Dobler}(2002)}]{brandenburg&dobler02}
{Brandenburg}, A. \& {Dobler}, W. 2002, Computer Physics Communications, 147,
  471

\bibitem[{{Brandenburg} {et~al.}(1995){Brandenburg}, {Nordlund}, {Stein}, \&
  {Torkelsson}}]{brandenburgetal95}
{Brandenburg}, A., {Nordlund}, A., {Stein}, R.~F., \& {Torkelsson}, U. 1995,
  \apj, 446, 741

\bibitem[{{Cargo} \& {Gallice}(1997)}]{cargo&gallice97}
{Cargo}, P. \& {Gallice}, G. 1997, Journal of Computational Physics, 136, 446

\bibitem[{{Chandrasekhar}(1961)}]{chandrabook61}
{Chandrasekhar}, S. 1961, {Hydrodynamic and hydromagnetic stability}
  (International Series of Monographs on Physics, Oxford: Clarendon, 1961)

\bibitem[{{Crockett} {et~al.}(2005){Crockett}, {Colella}, {Fisher}, {Klein}, \&
  {McKee}}]{Crockett05}
{Crockett}, R.~K., {Colella}, P., {Fisher}, R.~T., {Klein}, R.~I., \& {McKee},
  C.~F. 2005, Journal of Computational Physics, 203, 422

\bibitem[{{Dedner} {et~al.}(2002){Dedner}, {Kemm}, {Kr{\" o}ner}, {Munz},
  {Schnitzer}, \& {Wesenberg}}]{Dedner02}
{Dedner}, A., {Kemm}, F., {Kr{\" o}ner}, D., {et~al.} 2002, Journal of
  Computational Physics, 175, 645

\bibitem[{Evans \& Hawley(1988)}]{evans&hawley88}
Evans, C. \& Hawley, J. 1988, ApJ, 33, 659

\bibitem[{{Falle}(2002)}]{falle02}
{Falle}, S.~A.~E.~G. 2002, \apjl, 577, L123

\bibitem[{{Gardiner} \& {Stone}(2005{\natexlab{a}})}]{gardiner&stone05a}
{Gardiner}, T.~A. \& {Stone}, J.~M. 2005{\natexlab{a}}, Journal of
  Computational Physics, 205, 509

\bibitem[{{Gardiner} \& {Stone}(2005{\natexlab{b}})}]{gardiner&stone05b}
{Gardiner}, T.~A. \& {Stone}, J.~M. 2005{\natexlab{b}}, ArXiv Astrophysics
  e-prints

\bibitem[{{Goldreich} \& {Lynden-Bell}(1965)}]{goldreich&lyndenbell65}
{Goldreich}, P. \& {Lynden-Bell}, D. 1965, MNRAS, 130, 125

\bibitem[{{Hawley} {et~al.}(1995){Hawley}, {Gammie}, \&
  {Balbus}}]{hawleyetal95}
{Hawley}, J.~F., {Gammie}, C.~F., \& {Balbus}, S.~A. 1995, ApJ, 440, 742

\bibitem[{{Hennebelle} {et~al.}(2004){Hennebelle}, {Whitworth}, {Cha}, \&
  {Goodwin}}]{hennebelle04}
{Hennebelle}, P., {Whitworth}, A.~P., {Cha}, S.-H., \& {Goodwin}, S.~P. 2004,
  \mnras, 348, 687

\bibitem[{{Hosking} \& {Whitworth}(2004)}]{hosking&whitworth04}
{Hosking}, J.~G. \& {Whitworth}, A.~P. 2004, \mnras, 347, 1001

\bibitem[{Khokhlov(1998)}]{Khokhlov98}
Khokhlov, A.~M. 1998, J. Comput. Phys., 143, 519

\bibitem[{{Londrillo} \& {Del Zanna}(2000)}]{londrillo&delzanna00}
{Londrillo}, P. \& {Del Zanna}, L. 2000, \apj, 530, 508

\bibitem[{{Londrillo} \& {Del Zanna}(2004)}]{londrillo&delzanna04}
{Londrillo}, P. \& {Del Zanna}, L. 2004, Journal of Computational Physics, 195,
  17

\bibitem[{{Matsumoto} \& {Hanawa}(2003)}]{matsumoto&hanawa03}
{Matsumoto}, T. \& {Hanawa}, T. 2003, \apj, 595, 913

\bibitem[{{Miyoshi} \& {Kusano}(2005)}]{miyoshi&kusano05}
{Miyoshi}, T. \& {Kusano}, K. 2005, Journal of Computational Physics, 208, 315

\bibitem[{{Papaloizou} {et~al.}(2004){Papaloizou}, {Nelson}, \&
  {Snellgrove}}]{papaloizouetal04}
{Papaloizou}, J.~C.~B., {Nelson}, R.~P., \& {Snellgrove}, M.~D. 2004, MNRAS,
  350, 829

\bibitem[{{Phillips} \& {Monaghan}(1985)}]{phillips&monaghan85}
{Phillips}, G.~J. \& {Monaghan}, J.~J. 1985, MNRAS, 216, 883

\bibitem[{{Powell} {et~al.}(1999){Powell}, {Roe}, {Linde}, {Gombosi}, \& {de
  Zeeuw}}]{powell99}
{Powell}, K.~G., {Roe}, P.~L., {Linde}, T.~J., {Gombosi}, T.~I., \& {de Zeeuw},
  D.~L. 1999, Journal of Computational Physics, 154, 284

\bibitem[{{Price} \& {Monaghan}(2004{\natexlab{a}})}]{price&monaghan04a}
{Price}, D.~J. \& {Monaghan}, J.~J. 2004{\natexlab{a}}, \mnras, 348, 123

\bibitem[{{Price} \& {Monaghan}(2004{\natexlab{b}})}]{price&monaghan04b}
{Price}, D.~J. \& {Monaghan}, J.~J. 2004{\natexlab{b}}, \mnras, 348, 139

\bibitem[{{Ryu} \& {Jones}(1995)}]{ryu&jones95}
{Ryu}, D. \& {Jones}, T.~W. 1995, \apj, 442, 228

\bibitem[{{Ryu} {et~al.}(1998){Ryu}, {Miniati}, {Jones}, \& {Frank}}]{ryu98}
{Ryu}, D., {Miniati}, F., {Jones}, T.~W., \& {Frank}, A. 1998, \apj, 509, 244

\bibitem[{{Stone} \& {Norman}(1992{\natexlab{a}})}]{stone&norman92a}
{Stone}, J.~M. \& {Norman}, M.~L. 1992{\natexlab{a}}, ApJS, 80, 753

\bibitem[{{Stone} \& {Norman}(1992{\natexlab{b}})}]{stone&norman92b}
{Stone}, J.~M. \& {Norman}, M.~L. 1992{\natexlab{b}}, ApJS, 80, 791

\bibitem[{{T{\' o}th}(2000)}]{toth00}
{T{\' o}th}, G. 2000, Journal of Computational Physics, 161, 605

\bibitem[{{Teyssier}(2002)}]{teyssier02}
{Teyssier}, R. 2002, A\&A, 385, 337

\bibitem[{{Teyssier} {et~al.}(2006){Teyssier}, {Fromang}, \&
  {Dormy}}]{teyssieretal06}
{Teyssier}, R., {Fromang}, S., \& {Dormy}, E. 2006, {Journal of Computational
  Physics}, submitted

\bibitem[{Toro(1997)}]{toro97}
Toro, E. 1997, Riemann solvers and numerical methods for fluid dynamics
  (Springer)

\bibitem[{{Torrilhon}(2004)}]{torrilhon04}
{Torrilhon}, M. 2004, Journal of Plasma Physics, 69, 253

\bibitem[{{Torrilhon} \& {Balsara}(2004)}]{torrilhon&balsara04}
{Torrilhon}, M. \& {Balsara}, D.~S. 2004, Journal of Computational Physics,
  201, 586

\bibitem[{T{\'o}th \& Roe(2002)}]{toth02}
T{\'o}th, G. \& Roe, P.~L. 2002, J. Comput. Phys., 180, 736

\bibitem[{{van Leer}(1977)}]{vanleer77}
{van Leer}, B. 1977, Journal of Computational Physics, 23, 276

\bibitem[{{Velikhov}(1959)}]{velikhov59}
{Velikhov}, E.~P. 1959, J.Exp.Theor.Phys.(USSR), 36, 1398

\bibitem[{{Ziegler}(2004)}]{ziegler04}
{Ziegler}, U. 2004, Journal of Computational Physics, 196, 393

\bibitem[{{Ziegler}(2005)}]{ziegler05}
{Ziegler}, U. 2005, \aap, 435, 385

\bibitem[{{Ziegler} \& {Yorke}(1997)}]{ziegler&yorke97}
{Ziegler}, U. \& {Yorke}, H.~W. 1997, Computer Physics Communications, 101, 54

\end{thebibliography}

\end{document}